%% file: main.tex
\documentclass[sigconf,nonacm]{acmart}

\usepackage{algorithmic}
\usepackage{graphicx}
\usepackage{textcomp}
\usepackage{xcolor}
\usepackage[normalem]{ulem}
\usepackage{bbding}
\usepackage{enumitem}
\usepackage{pifont}
\usepackage[linesnumbered, ruled, vlined]{algorithm2e}
\usepackage[makeroom]{cancel}
\usepackage{extarrows}
\usepackage{multirow}
\usepackage{booktabs}
\usepackage{tabularx,framed}
\usepackage{makecell}
\usepackage{courier}

\AtBeginDocument{%
  }

\begin{document}

\input{defines}

\title[]{A Full-Stack Performance Evaluation Infrastructure for 3D-DRAM-based LLM Accelerators}

\author{Cong Li\textsuperscript{1} \hspace{0.5em} 
Chenhao Xue\textsuperscript{1} \hspace{0.5em} Yi Ren\textsuperscript{1} \hspace{0.5em}
Xiping Dong\textsuperscript{1} \hspace{0.5em} Yu Cheng\textsuperscript{1} \hspace{0.5em}
Yinbo Hu\textsuperscript{1} \hspace{0.5em}
Fujun Bai\textsuperscript{2} \hspace{0.5em} \\
Yixin Guo\textsuperscript{2} \hspace{0.5em} Xiping Jiang\textsuperscript{2} \hspace{0.5em}
Qiang Wu\textsuperscript{3} \hspace{0.5em}
Zhi Yang\textsuperscript{1} \hspace{0.5em}
Zhe Cheng\textsuperscript{1} \hspace{0.5em}
Yuan Xie\textsuperscript{4} \hspace{0.5em}
Guangyu Sun\textsuperscript{1}\\
\normalsize{\textsuperscript{1}\textit{Peking University} \quad
\textsuperscript{2}\textit{Xi'an UniIC Semiconductors} \quad
\textsuperscript{3}\textit{Houmo AI} \quad
\textsuperscript{4}\textit{HKUST}}}
\renewcommand{\shortauthors}{}

%%%%%% -- PAPER CONTENT STARTS-- %%%%%%%%

\input{tex/0-abstract}

\maketitle

\input{tex/1-introduction}

\input{tex/2-background}

\input{tex/3-simulator}

\input{tex/4-cloud-chip-dse}

\input{tex/5-edge-chip-dse}
\input{tex/8-conclusion}

\bibliographystyle{plain}
\bibliography{refs}

\end{document}

%% file: defines.tex
\newcommand{\squishlist}{
 \begin{list}{$\bullet$}
  { \setlength{\itemsep}{0pt}
     \setlength{\parsep}{2.25pt}
    \setlength{\topsep}{0pt}
     \setlength{\partopsep}{0pt}
     \setlength{\leftmargin}{1.5em}
     \setlength{\labelwidth}{1em}
     \setlength{\labelsep}{0.5em} } }
\newcommand{\squishend}{
  \end{list}  }

\newcommand{\bone}{\ding{182}}
\newcommand{\btwo}{\ding{183}}
\newcommand{\bthree}{\ding{184}}
\newcommand{\bfour}{\ding{185}}
\newcommand{\bfive}{\ding{186}}
\newcommand{\bsix}{\ding{187}}
\newcommand{\bseven}{\ding{188}}
\newcommand{\beight}{\ding{189}}
\newcommand{\bnine}{\ding{190}}
\newcommand{\bten}{\ding{191}}
\newcommand{\beleven}{\ding{192}}

\newcommand{\todo}[1]{\textcolor{blue}{#1}}

\definecolor{interfacecolor}{HTML}{282FAA}
\newcommand{\interface}[1]{\textcolor{interfacecolor}{#1}}

\newcommand{\cell}[1]{\begin{tabular}[c]{@{}c@{}}#1\end{tabular}}

\newcommand{\TitleAbbr}{ATLAS}

\newcommand{\placeholder}{{\color{gray}{xxx xx xx x xx x xx xxxx xxxx xxx xx xx x xx x xx xxxx xxxxxxx xx xx x xx x xx xxxx xxxx xxx xx xx x xx x xx xxxx xxxx xxx xx xx x xx x xx xxxx xxxx xxx xx xx x xx x xx xxxx xxxx xxx xx xx x xx x xx xxxx xxxx xxx xx xx x xx x xx xxxx xxxx xxx xx xx x xx x xx xxxx xxxx xxx xx xx x xx x xx xxxx xxxx xxx xx xx x xx x xx xxxx xxxx xxx xx xx x xx x xx xxxx xxxx xxx xx xx x xx x xx xxxx xxxx xxx xx xx x xx x xx xxxx xxxx. }}}

\newcommand{\placeholdershort}{{\color{gray}{xxx xx xx x xx x xx xxxx xxxx xxx xx xx x xx x xx xxxx xxxxxxx xx xx x xx x xx xxxx xxxx xxx xx xx x xx x xx xxxx xxxx xxx xxxx xxxx xxx xx xx x xx x xx xxxx xxxx xxx xx xx x xx x xx xxxx xxxx. }}}

%% file: tex/0-abstract.tex
\begin{abstract}

Large language models (LLMs) exhibit memory-intensive behavior during decoding, 
making it a key bottleneck in LLM inference.
To accelerate decoding execution, hybrid-bonding-based 3D-DRAM has been adopted in LLM accelerators.
While this emerging technology provides strong performance gains over existing hardware, current 3D-DRAM accelerators (3D-Accelerators) rely on closed-source evaluation tools, limiting access to publicly available performance analysis methods.
Moreover, existing designs are highly customized for specific scenarios, lacking a general and reusable full-stack modeling for 3D-Accelerators across diverse usecases.

To bridge this fundamental gap, we present \TitleAbbr, the first silicon-proven \uline{A}rchitectural \uline{T}hree-dimesional-DRAM-based \uline{L}LM \uline{A}ccelerator \uline{S}imulation framework.
Built on commercially deployed multi-layer 3D-DRAM technology, \TitleAbbr~introduces unified abstractions for both 3D-Accelerator architecture and programming primitives to support arbitrary LLM inference scenarios.
Validation against real silicon shows that \TitleAbbr~achieves $\le$8.57\% simulation error and 97.26-99.96\% correlation with measured performance.
Through design space exploration with \TitleAbbr, we demonstrate its ability to guide architecture design and distill key takeaways for both 3D-DRAM memory system and 3D-Accelerator microarchitecture across scenarios.
\TitleAbbr~will be open-sourced upon publication,
enabling further research on 3D-Accelerators.

\end{abstract}

%% file: tex/1-introduction.tex
\section{Introduction}

Large language models (LLMs) have achieved remarkable performance on generative tasks such as code generation~\cite{chen2021evaluating,nijkamp2022codegen}, problem reasoning~\cite{wei2022emergent,wei2022chain}, and personal assistants~\cite{mei2024aios,huang2022inner}.
This capability has led to their widespread deployment in scenarios such as datacenter services~\cite{chatgpt,gemini,claude,deepseek} or personalized edge devices~\cite{jetson-chat,durante2024agent,huang2022language,hensel2023large}.
Efficient LLM inference requires accelerating its two stages: prefill and decoding. 
In the prefill stage, the LLM processes the entire input sequence (prompt) to generate the first output token,
while in decoding it generates one token at a time using the previous token as input, autoregressively constructing the output sequence.

In LLM inference, prefill is compute-intensive due to high arithmetic intensity from processing long prompts in parallel,
making it well-suited for high-throughput xPUs (e.g., GPUs/TPUs).
In contrast, decoding proceeds token by token, exposing much lower  parallelism and shifting the bottleneck to memory access.
Although HBM (in datacenter) or LPDDR-DRAM (at edge) are used to boost memory bandwidth, this remains insufficient to match xPUs' massive compute capability (up to several TB/s vs. hundreds of TFLOPS), making decoding the primary bottleneck in LLM inference.

As a promising alternative to conventional memory systems, hybrid-bonding-based 3D-DRAM has recently been adopted in LLM accelerators~\cite{li2025h2,pan2025stratum,cao20261}.
Compared to HBM/LPDDR-DRAM, it offers higher bandwidth and lower energy through increased I/O density and shorter data paths.
It can also customize compute logic in CMOS technology on the separate logic die, ensuring sufficient compute capability.
As a result, in both cloud~\cite{pan2025stratum} and edge~\cite{li2025h2,cao20261} settings, 3D-DRAM-based accelerators (3D-Accelerators) achieve substantial decoding speedups over conventional xPU-based systems and near-memory processing designs that integrate compute units into traditional memory modules~\cite{heo2024neupims,kim2024sk,kim2023samsung,li2024specpim,park2024attacc,yun2024duplex,he2025papi,gu2025pim,lee2021hardware,he2020newton,kim2024breakthrough,lee20221ynm}.

Despite these advantages, there is still a lack of systematic softwa-re-hardware modeling and performance evaluation 
methodologies for 3D-Accelerators, limiting broader architecture exploration:
First, existing designs lack detailed modeling of both the 3D-DRAM memory system and the 3D-Accelerator compute architecture~\cite{li2025h2,cao20261,pan2025stratum}.
A general architecture abstraction applicable across diverse scenarios is still missing.
Second, existing designs customize execution dataflow for specific models (e.g., mixture-of-experts~\cite{pan2025stratum}) or scenarios (e.g., edge inference~\cite{li2025h2,cao20261}).
They lack a unified design for programming model and primitives, limiting flexible expression of software execution.
Third, existing designs rely on closed-source test chips~\cite{cao20261} or simulators~\cite{li2025h2,pan2025stratum}, leaving no publicly available and accurate tool for 3D-Accelerator performance evaluation.

To bridge these fundamental gaps, we propose \TitleAbbr, the first silicon-proven full-stack 3D-Accelerator simulation framework.
We begin by modeling the 3D-DRAM memory system based on commercially deployed manufacturing technology,
and construct a general 3D-Accelerator architecture template that captures 3D-DRAM organization properties.
On top of this template, we design a hierarchical programming model and corresponding primitives guided by industry best practices to enable flexible operator customization.
Based on these unified abstractions, we implement \TitleAbbr~to enable automated translation from architecture parameters and operator implementations to cycle-level performance.
After verifying \TitleAbbr's fidelity against real silicon measurements, we demonstrate how it guides architecture design and distill key takeaways for both the 3D-DRAM memory system and 3D-Accelerator microarchitecture.
To summarize, we have made the following contributions:
\squishlist
    \item We propose a general 3D-Accelerator architecture template based on mature 3D-DRAM manufacturing technology.
    \item We develop 3D-Accelerator programming model and primitives by incorporating industry deployment best practices.
    \item We build \TitleAbbr, the first full-stack 3D-Accelerator 
    simulation framework, and validate it against real silicon measurements.
    \item We leverage \TitleAbbr~to guide 3D-Accelerator design space exploration and extract key takeaways for future architectures.
\squishend
Experiments show that \TitleAbbr~achieves $\le$8.57\% error and 97.26-99.96\% correlation with real hardware, while its derived designs deliver up to 3.64/1.42× speedup over GPUs and prior 3D-Accelerators.

%% file: tex/2-background.tex
\section{Background and Motivation}

\subsection{Transfomer-based Large Language Models}

As shown in Fig.~\ref{fig:llm}, LLMs are built on decoder transformer layers, each including an attention block, a feed-forward network (FFN) block, along with normalization and residual operators~\cite{vaswani2017attention}.
In the attention block, each token is first projected by a fully connected (FC) layer to produce query/key/value (Q/K/V) vectors,
which are processed by self-attention and another FC layer to generate the output.
In self-attention, Q/K/V vectors are partitioned into multiple heads, where each Q interacts with corresponding KV vectors.
To avoid re-computation, KV vectors associated with each token are persisted and reused throughout inference, which is known as KV Cache~\cite{pope2023efficiently}.
Each KV head serves one or multiple Q heads, corresponding to multi-head attention (MHA) and group-query attention (GQA)~\cite{ainslie2023gqa}, respectively.
In the FFN block, dense LLMs adopt multi-layer perceptron (MLP) or gated linear unit (GLU)~\cite{shazeer2020glu},
where inputs pass through bottom FCs (FC1/FC3), activation function, and a top FC (FC2) to produce the output.
In mixture-of-experts (MoE) models, the FFN block contains multiple expert FFNs.
For each token, a gating network selects top-K experts for computation, and their outputs are then combined with weighted aggregation.

\input{fig_tex/llm}

\input{fig_tex/hybrid-bonding}

\subsection{Hybrid-Bonding-based 3D-DRAM}

Hybrid bonding (HB) is a next-generation integration technology emerged in recent years~\cite{fujun2020stacked,niu2022184qps,yue2024exploiting,wang2023135,wang20253d,li2025h2,pan2025stratum,cao20261}.
As shown in Fig.~\ref{fig:hybrid-bonding}-(a), HB 3D-stacks the DRAM die on top of the logic die and connects them with copper pillars.
It significantly increases I/O density (110000/mm$^2$, 3$\mu$m pitch), bringing an order-of-magnitude bandwidth density improvement against HBM~\cite{wang2023135}.
At the same time, the shortened data path reduces memory access energy to 0.66-0.88pJ/bit, corresponding to a 77\%-83\% reduction over HBM~\cite{wang2023135}.
In addition, the logic die supports customized CMOS compute logic, ensuring sufficient compute capability alongside the increased bandwidth.
With the technlogy advancement, HB also supports multi-layer DRAM stacking.
As shown in Fig.~\ref{fig:hybrid-bonding}-(b), adjacent DRAM dies are connected by mini-TSVs to provide independent I/O datapaths, simultaneously improving memory capacity and bandwidth.

HB's high bandwidth aligns well with the memory-intensive nature of LLM decoding, motivating recent 3D-Accelerator proposals for LLM inference~\cite{li2025h2,pan2025stratum,cao20261}.
In edge scenarios, H$^2$-LLM~\cite{li2025h2} integrates 3D-Accelerators into NPU memory system via LPDDR5 interfaces, enabling end-to-end LLM acceleration.
Cao et al.~\cite{cao20261} extends this design by stacking two DRAM dies and using a more advanced logic node (40nm to 28nm).
In cloud settings, Stratum~\cite{pan2025stratum} stacks monolithic DRAM onto a logic die using single-DRAM-die bonding in Fig.~\ref{fig:hybrid-bonding}-(a), accelerating decoding for MoE models.

\subsection{Missing Foundation in 3D-DRAM Research}

Despite 3D-Accelerators' significant performance gains, there is still no general and faithful methodology to evaluate their behavior.
The key challenge is to jointly establish (1) a general and configurable full-stack abstraction for performance modeling, and (2) a reliable and abstraction-aligned infrastructure for performance evaluation.
Existing works still fall short in the following aspects:

\uline{\textit{Gap1: Lack of general and explorable architecture modeling.}}
First, existing works lack detailed and flexible 3D-DRAM modeling.
H$^2$-LLM~\cite{li2025h2} only models 3D-DRAM bandwidth provision,
without internal organization.
Stratum and Cao et al.~\cite{pan2025stratum,cao20261} only provide fixed DRAM array structures, preventing exploration of DRAM organization impact on bandwidth utilization.
Second, existing works adopt fixed or partially abstracted 3D-Accelerator architectures.
Stratum and Cao et al.~\cite{pan2025stratum,cao20261} empirically allocate compute, memory, and interconnect resources, without modeling the design space.
H$^2$-LLM~\cite{li2025h2} explores bandwidth and SRAM provision under a restricted architecture template without vector units and interconnect, limiting both generality and achievable performance.

\uline{\textit{Gap2: Lack of programmable and reusable software abstraction.}}
Existing works use workload-specific, hard-coded execution flows tightly coupled to target hardware, preventing their reuse across 3D-Accelerator configurations.
They also lack a unified programming interface, preventing flexible operator customization.
Meanwhile, tile-based domain-specific languages (DSLs) for deep learning operators (e.g., Triton, TileLang, etc.)~\cite{wang2025tilelang,tillet2019triton,cutile,iris} have become mainstream programming interfaces in modern accelerators due to their high usability.
However, existing 3D-Accelerators lack compatible programming abstractions, thus preventing their integration with these DSLs and limiting their usability in practical deployment.

\uline{\textit{Gap3: Lack of evaluation infrastructure.}}
Existing works rely on closed-source test chips~\cite{cao20261} or in-house simulators~\cite{li2025h2,pan2025stratum}, with no available framework for accurate 3D-Accelerator evaluation.
Moreover, due to the absence of full-stack modeling addressing \uline{\textit{Gap1-2}}, existing DRAM simulators~\cite{luo2023ramulator,li2020dramsim3} and LLM accelerator simulators~\cite{zhang2024llmcompass,duplex-sim} cannot be directly applied to 3D-Accelerator evaluation.
In addition, thermal effects must be considered, as stacked DRAM dies are directly affected by compute-induced heat, which may degrade DRAM retention.
While existing thermal simulator~\cite{han20222} can model transient temperature, they lack accurate modeling of material properties specific to 3D integration 
(HB interfaces, TSVs, BEOL layers, etc.), preventing direct thermal modeling of 3D-Accelerators.

%% file: fig_tex/llm.tex
\begin{figure}
    \centering
    \includegraphics[width=0.99\columnwidth]{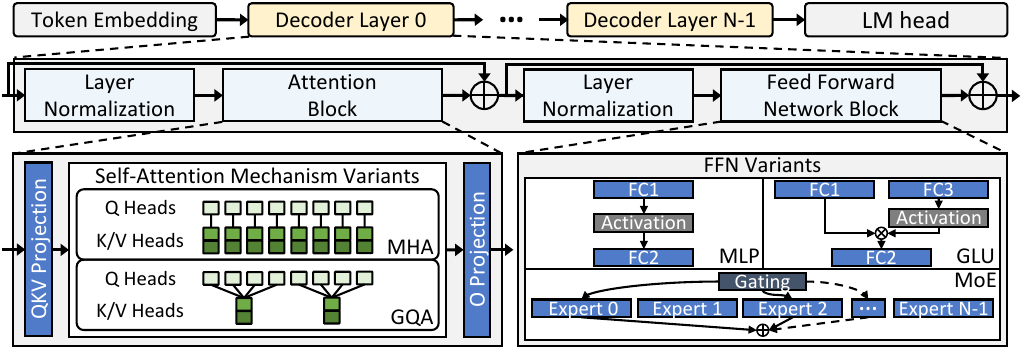}
    \caption{Structure of Transformer-based LLMs.}
    \label{fig:llm}
\end{figure}

%% file: fig_tex/hybrid-bonding.tex
\begin{figure}
    \centering
    \includegraphics[width=0.99\columnwidth]{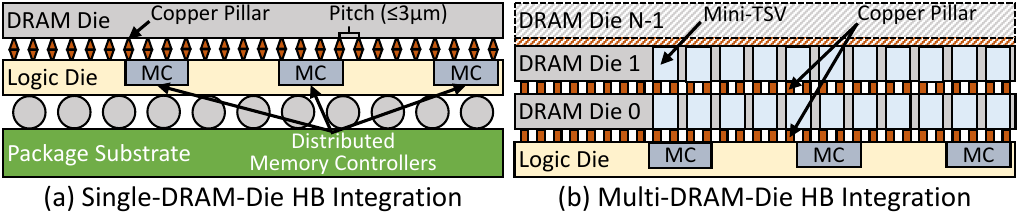}
    \caption{Overview of Hybrid-Bonding-based 3D-DRAM.}
    \label{fig:hybrid-bonding}
\end{figure}

%% file: tex/3-simulator.tex
\section{\TitleAbbr}

To bridge these gaps, we propose \TitleAbbr, the first full-stack evaluation framework for 3D-Accelerators.
In this section, building upon mature industrial 3D-DRAM manufacturing technologies, we first abstract the architectures of 3D-DRAM and 3D-Accelerator to establish a generic hardware template.
Then, we present programming model and primitive interfaces for 3D-Accelerator, illustrating their usage in core LLM inference operators.
Based on these abstractions, we introduce \TitleAbbr~with its structure, workflow, and extensibility, and validate its fidelity
using a 3D-Accelerator silicon implementation provided by our industry collaborators.

\input{fig_tex/3d-dram}

\subsection{Hardware Abstraction for 3D-Accelerators}
\label{sec:hardware-abstraction}

\noindent\textbf{3D-DRAM Architecture:}
As shown in Fig.~\ref{fig:3d-dram}-(a), in hybrid-bonding-based 3D-DRAM, each DRAM die comprises multiple homogeneous physical banks (PBs), serving as the basic unit of DRAM arrays.
PBs are grouped into memory channels, whose architecture is depicted in Fig.~\ref{fig:3d-dram}-(b).
In each channel, PBs are arranged as a 2D-array, collectively forming a single logical bank (LB).
The $R$ PB rows scale the memory capacity.
In each PB row, DRAM rows of $C$ PBs are concatenated into a logical row, serving as the basic unit for activate/precharge operations (ACT/PRE).
With mini-TSVs directly delivering power to PBs, simultaneous PB ACT/PRE is supported~\cite{jiang2024ssa}, ensuring the data supply for massive HB-I/O pins.
All PBs share the same I/O bus, and each DRAM access involves only one PB.

With multiple stacked DRAM dies, LBs can be organized flexibly.
As shown in Fig.~\ref{fig:3d-dram}-(c), consider two dies, each containing four PBs. 
Under LB configuration of $R$=$C$=2, the four PBs on a single die can be grouped into one LB, or two physically aligned PBs across the two dies can be grouped instead.
In practice, manufacturers select the organization with the shortest data path to reduce memory access energy. 
Under mature technology, inter-die timing skew in 3D-DRAM is below 50ps~\cite{wang20253d}. Therefore, with a fixed PB configuration, changing LB organization does not affect DRAM timing.

\noindent\textbf{3D-Accelerator Architecture:}
HB-I/O's copper pillars pass vertically through the DRAM stack to the logic die.
Given the 2D PB layout in DRAM dies, memory controllers (MCs) are correspondingly distributed across the logic die.
To align with such organization,
3D-Accelerators adopt a distributed multi-core architecture. As shown in Fig.~\ref{fig:chip}-(a), PBs on each DRAM die are grouped into bank partitions of equal size.
Vertically aligned partitions across DRAM dies are assigned to the corresponding compute core with comparable area budgets.
Each core directly accesses its aligned partitions, which are organized into memory channels shown in Fig.~\ref{fig:3d-dram}-(b).
Inter-core communication is handled by a dedicated interconnect under a non-uniform memory access (NUMA) abstraction.

Building on this NUMA property, the architecture abstraction of a multi-chip 3D-Accelerator system is depicted in Fig.~\ref{fig:chip}-(b).
Each chip contains multiple homogeneous cores. In each core, a 3D-DRAM memory system is coupled with compute units, including a matrix engine, a vector engine, and an SRAM buffer.
Each core also includes a router interfacing with the core-to-core interconnect for inter-core communication.
A core controller coordinates the execution of intra-core components and interacts with the chip controller to synchronize execution progress.
The chip controller globally manages operator execution progress, and communicates with other chips through the chip-to-chip interconnect.
Existing LLM-inference 3D-Accelerators~\cite{li2025h2,pan2025stratum,cao20261} are integrated into xPU memory systems, where inter-chip communication is handled via xPU accesses to their external memory interfaces.
Alternatively, they can also operate as standalone processors,
where communication is performed through scale-up interconnects (e.g., NVLink, CXL, etc.)~\cite{ualink,cxl-doc,nvlink} or scale-out networks (e.g., RDMA)~\cite{infiniband,gangidi2024rdma}.

\input{fig_tex/chip}

\input{table_tex/architecture-template}

\noindent\textbf{Architecture Description Template:}
Table~\ref{tab:3d-accelerator-system} summarizes key parameters to describe the 3D-Accelerator system.
Each core is parameterized by its memory system and compute resources.
For the memory system, each PB is specified by its row size and row count.
A LB is configured by PB count along the row ($R$) and column ($C$) dimensions.
The bandwidth of each channel is determined by the I/O pin count and per-pin data rate.
With the channel count specified, each core's total memory bandwidth and capacity can be derived.
For the compute logic, the processing capability is described by the compute throughput of the matrix and vector engines, together with the capacity and bandwidth of the SRAM buffer.

The inter-core interconnect is modeled as a Network-on-Chip (NoC).
Accordingly, we specify parameters related to link latency (e.g., router pipeline delay), link bandwidth (e.g., flit size, link width, etc.), as well as the NoC topology, to evaluate communication performance.
For inter-accelerator communication, to unify both memory interface and scale-up/out interconnects, we follow the practice in existing simulators~\cite{hyun2024pathfinding,zhang2024llmcompass} and adopt a fixed-bandwidth model (i.e., communication\_latency = link\_latency + transfer\_size / interconnect\_bandwidth) 
to evaluate its performance.

\subsection{Programming 3D-Accelerator Systems}

\noindent\textbf{Programming Model:}
Under
the above architecture abstraction, programming the 3D-Accelerator system can be viewed across four levels from a bottom-up perspective:
In each accelerator, programmers need to manage \textit{\uline{(1) computation in each core}} as well as \textit{\uline{(2) communication among cores}}.
Across multiple accelerators, they need to coordinate \textit{\uline{(3) computation in each accelerator}} and \textit{\uline{(4) communication among accelerators}}.
Guided by the characteristics of LLM inference workloads and their deployment patterns, we adopt the following programming models for these four levels:

\textit{\uline{(1) Core-Level Execution}}:
LLM operators involve huge tensors with dimensions in the thousands.
In practical deployments, all cores in an accelerator execute the same operator simultaneously and process the operator graph sequentially.
Accordingly, intra-accelerator computation follows the single-program multiple-data (SPMD) model.
It simplifies programming and makes load balancing straightforward, as we only need to maintain identical tensor shapes across cores during operator partition.
It also aligns with existing multi-core accelerators with uniform memory abstraction (e.g. GPUs), improving the deployability of 3D-Accelerators.

\textit{\uline{(2) Core-Level Communication}}: % MPMD
3D-Accelerator's NUMA nature requires explicit inter-core communication:
After all cores complete operator computation, partial sums need to be accumulated and redistributed across cores to form the next operator's input layout.
To accommodate diverse interconnect topologies and flexible operator partition, we adopt the multiple-program multiple-data (MPMD) programming model, allowing each core to exchange data with different peers under arbitrary communication patterns.

\textit{\uline{(3) Multi-Accelerator Execution}}:
When deploying LLMs across multiple accelerators, industrial best practices adopt model parallelism, where model weights are evenly partitioned across devices~\cite{deepseekai2024deepseekv3technicalreport,qin2025mooncake,kwon2023efficient,sglang}.
Consequently, accelerators execute the same program, allowing us to adopt the SPMD programming model.

\textit{\uline{(4) Multi-Accelerator Communication}}:
Under model parallelism, cross-accelerator communication is implemented via symmetric collective primitives.
Accordingly, we can adopt the same SPMD model as mainstream collective communication libraries~\cite{nccl,torch-distributed}.

\input{table_tex/programming-interface}

\noindent\textbf{Primitive Abstraction:}
Since multi-accelerator execution shares the same program under the SPMD model, and multi-accelerator communication can reuse existing collective primitives, we focus programming interface design
on core-level execution and communication, jointly supporting SPMD and MPMD to accommodate the NUMA nature of 3D-Accelerators.
Table~\ref{tab:programming-interface} summarizes a partial list of key programming primitives.
To improve usability, we implement these interfaces in the Python ecosystem and align their semantics with mainstream operator programming DSLs~\cite{tillet2019triton,wang2025tilelang}.
Core-level execution is supported by the following interfaces:

\textit{\uline{(1) DRAM tensor declaration.}} 
The \texttt{\interface{tensor}} interface defines the data layout in each core's 3D-DRAM system. Since cores follow the SPMD model, the same DRAM layout can be applied to all cores.

\textit{\uline{(2) SRAM tile declaration.}} 
The \texttt{\interface{alloc}} interface specifies the shape of each data tile stored in the SRAM buffer. The total size of all SRAM tiles is constrained by the SRAM capacity.

\textit{\uline{(3) Data transfer.}} 
To simplify programming, all data movement is expressed by a unified \texttt{\interface{copy}} interface. 
When \texttt{src} is defined by \texttt{\interface{tensor}} and \texttt{dst} by \texttt{\interface{alloc}}, data are loaded from DRAM to SRAM. 
Conversely, data are written back to DRAM.
When both are defined by \texttt{\interface{alloc}}, the operation represents an SRAM buffer copy.

\textit{\uline{(4) Computation.}} 
For tensor computation, we provide \texttt{\interface{gemm}} interface, as GEMM can express all mainstream tensor workloads.
For vector workloads, reduction/elementwise interfaces are provided.

Core-level communication contains the following interfaces:

\textit{\uline{(1) Core organization declaration.}}
\texttt{\interface{core\_array}} logically organizes each accelerator's all cores into an N-dimensional array for operator partition.
Based on the physical core layout shown in Fig.~\ref{fig:chip}-(a), the mapping between logical coordinates and 2D physical coordinates
is defined as follows:
Suppose \texttt{shape}=($X_0,...,X_{N-1}$), where $X \cdot Y$=$\prod_{i=0}^{N-1} X_i$.
For a core with logical coordinate ($x_0,...,x_{N-1}$), let $c$=$\sum_{i=0}^{N-1} x_i \cdot X_i$.
Its physical coordinate ($m,n$) is computed as: $x$=$\lfloor \frac{c}{Y} \rfloor$, $y$=$c \bmod Y$.
\texttt{shape}'s definition is independent of the inter-core interconnect topology. 
However, the topology should be considered in operator partition design to maximize performance.

\textit{\uline{(2) Operator partition.}}
\texttt{\interface{split\_gemm}} describes the partition of LLM FC operators.
Given operator shape
($M$,$K$)×($K$,$N$), \texttt{core\_dim} \texttt{\_mapping} specifies which axes of \texttt{\interface{core\_array}} 
% are used to 
partition these dimensions.
Taking $N$ as an example, suppose it is partitioned along the $i_0,...,i_{t-1}$-th axes of \texttt{\interface{core\_array}}.
Each core then processes $N/\prod_{j=0}^{t-1} X_{i_j}$ elements, with shard offsets determined by its logical coordinate.
Shards are assigned contiguously from axis $i_{t-1}$ to $i_0$.
If $M$ is partitioned, matrix ($M$,$K$) is sharded, while matrix ($K$,$N$) is fully replicated accordingly.
If $K$ or $N$ is partitioned, matrix ($K,N$) is sharded, and ($M$,$K$) is replicated along the same $K$ partition.
These input placements uniquely determine the placement of the ($M$,$N$) output partial-sum shards.
For attention operators, the two GEMMs can be fused via online softmax~\cite{flashdecoding}.
We therefore provide \texttt{\interface{split\_attention}} to describe inter-core attention partition.
By specifying the token slot IDs in each core's KV cache tensor, it defines how attention computation is distributed across cores.

\textit{\uline{(3) Inter-core data transfer.}}
We provide flexible point-to-point primitives \texttt{\interface{send}}/\texttt{\interface{recv}} for inter-core communication. 
By combining these primitives, arbitrary communication patterns can be implemented.
Source and destination cores are specified using the logical coordinate defined earlier, with the linearized index $c$=$\sum_{i=0}^{N-1} x_i \cdot X_i$ passed to the \texttt{src}/\texttt{dst} arguments.
The primitives are topology-agnostic and support any core-pairs,
although topology-aware implementations are required for high performance.

\input{fig_tex/computation-programming}

\noindent\textbf{Programming Examples:}
Next, we demonstrate how the proposed primitives program 3D-Accelerators by implementing core LLM inference operators.
Fig.~\ref{fig:computation-programming} presents 
FC operator (\texttt{matmul}) and decoding fused attention (\texttt{fused\_attention}) examples.
Implementing a compute operator involves three steps: 
(1) Declare DRAM tensors.
Under the SPMD model, tensors are defined based on the per-core workloads. (2) Specify SRAM tiles, which decides the intra-core on-chip tiling strategy.
(3) Describe execution flow, including DRAM reads, on-chip execution, and DRAM writes.

For \texttt{matmul}, we first declare the matrices (\textbf{lines 2-4}), then define the computation tiles and corresponding SRAM buffers (\textbf{lines 5-8}).
Under output-stationary dataflow (\textbf{lines 9-11}), each iteration loads input tiles (\textbf{lines 12-13}), performs tile-level GEMM (\textbf{line 14}), and accumulates partial sums (\textbf{line 15}).
After the $K$ dimension is fully processed, the output tile is written back to DRAM (\textbf{line 16}).

For \texttt{fused\_attention}, we show the execution of one KV head.
Assume $N$ query heads share one KV head with dimension $H$ and context length $S$, and the tiling factor of $S$ is $tS$. 
We first load the query vector shared by all KV tiles (\textbf{line 14}).
For each KV tile, after preparing the DRAM and SRAM data (\textbf{lines 16-18}), we compute the query-key GEMM (\textbf{line 19}), convert the results into softmax scores via a sequence of vector operations (\textbf{lines 20-24}), and multiply them with value matrix (\textbf{line 25}).
Finally, partial sums are weight-accumulated (\textbf{lines 26-27}).
After processing the entire context, the final output is written back to DRAM (\textbf{line 28}).

\input{fig_tex/communication-programming}

Fig.~\ref{fig:communication-programming} illustrates how to implement inter-core operator partition and communication.
In Fig.~\ref{fig:communication-programming}-(a), eight cores are organized as a (2,4) array (\textbf{line 1}).
For GEMM partition, $N$ is split along the 1st axis of \texttt{\interface{core\_array}}, $K$ along the 0th axis, while $M$ is not partitioned (\textbf{line 6-9}).
Accordingly, the ($M$,$K$) matrix is split into two $K$-shards and assigned to cores (0,$j$) and (1,$j$) ($j$=0,1,2,3).
The ($K$,$N$) matrix is divided into eight shards and distributed across all cores following the core-axis order.
The output matrix ($M$,$N$) is partitioned into four shards, each consisting of two partial sums produced by the two cores sharing the same coordinate on the 1st axis.
For attention partition, assume a request with ten tokens.
Request tokens are consecutively distributed across cores following the item order in \texttt{token\_slot\_list}.
Each item specifies current core's logical coordinate and KV cache slot IDs to store the assigned tokens (detailed layout is described later).
Once the partition is defined, a single kernel invocation describes the computation across all cores.

For communication operators, Fig.~\ref{fig:communication-programming}-(b) shows a ring reduce-scatter example.
Since each core exchanges data with different peers, the communication kernel is implemented once (\textbf{lines 1-12}) but invoked for each core under the MPMD model (\textbf{lines 17-20}) to declare all communication operations.

\subsection{ATLAS Framework}

\noindent\textbf{Framework Overview:}
Based on the architecture and programming abstractions described above, we propose the \TitleAbbr~framework. As shown in Fig.~\ref{fig:atlas-framework}, \TitleAbbr~contains four modules: thermal analyzer, operator parser, intra-core tiling explorer, and cycle-level performance simulator.
The framework takes two inputs: (1) Architecture parameters, including both the architecture template (Table~\ref{tab:3d-accelerator-system}) and the power/area information required for thermal analysis. 
(2) LLM operator implementation, 
expressed by the programming primitives in Table~\ref{tab:programming-interface}.
For output, \TitleAbbr~produces the thermally feasible architecture and its inference performance.

\noindent\textbf{Evaluation Workflow:}
The workflow begins by parsing the operator implementation into an abstract syntax tree (AST), while regulating the architecture parameters with the thermal analyzer.
If the temperature exceeds the retention limit, the thermal analyzer progressively reduces the operating frequency until the constraint is satisfied, producing adjusted architecture parameters.
Then, the intra-core tiling explorer uses the adjusted architecture parameters and the primitive AST to derive operator tiling factors (e.g., $tM$,$tN$,$tK$,$tS$ in Fig.~\ref{fig:computation-programming}).
We provide an autotuning option in the programming interface: when enabled, the explorer searches for the best-performing configuration using the cycle-level performance simulator.
Otherwise, it uses the factors specified in the implementation.
After tuning, it generates the execution description and tensor placement for the performance simulator. 
Given the hardware configuration and the software inputs, the cycle-level simulator evaluates the inference performance.
The evaluation results and the thermally feasible architecture are finally reported.

\noindent\textbf{Simulator Software Inputs:}
Fig.~\ref{fig:atlas-framework} illustrates the formats of execution description and tensor placement using YAML syntax~\cite{yaml}.
The execution description represents the operator graph as an ordered list. For each entry, the \texttt{execution} field records per-iteration hardware workloads.
During generation, the tiling explorer automatically pipelines tile execution based on hardware dependencies.
For example, the GEMM operator in Fig.~\ref{fig:atlas-framework} requires matrices $A$ and $B$ to be loaded to the SRAM buffer before computation.
Accordingly, the first iteration in the \texttt{execution} field issues only DRAM reads for the initial tile. In the next iteration, computation on the loaded tile is issued together with DRAM reads for the next tile, enabling overlap between computation and memory access.

The tensor placement specifies each tensor’s DRAM layout.
During generation, the tiling explorer aligns base addresses with logical rows to maximize row-buffer utilization. 
If strides are not specified in operator implementation, tiling explorer automatically infers the optimal configuration based on the access pattern of each tensor.

\noindent\textbf{Simulator Organization:}
The simulator contains a 3D-Accelerator simulator and an inter-accelerator interconnect model. 
As discussed in Sec.~\ref{sec:hardware-abstraction}, for the interconnect model, we follow existing simulators' practice~\cite{hyun2024pathfinding,zhang2024llmcompass} and adopt the fixed bandwidth model to unify the modeling of memory interfaces and scale-up/out interconnects.
The 3D-Accelerator simulator includes a global manager, an inter-core interconnect simulator, and multiple core simulator 

\noindent{objects.} The global manager distributes simulation tasks to other modules at operator granularity.
It maintains a globally synchronized clock and advances execution after all iterations of the current operator have completed.
The inter-core interconnect simulator models 
cycle-level cross-core communication and exposes send/receive queues for each core simulator object to support full-duplex data transfer.

Each core simulator object contains four components: 
(1) Core scheduler. It receives execution description from the global manager, dispatches tasks to other intra-core modules, and reports operator completion to the global manager.
(2) Logic-die simulator.
It models the matrix engine, vector unit, and SRAM buffer to estimate on-chip computation latency. 
(3) 3D-DRAM simulator. It includes a front-end 
for low-level DRAM command generation 
and a cycle-accurate 3D-DRAM performance simulator.
(4) Interconnect wrapper. 
It interfaces with the interconnect simulator. 
The packet generator submits packets to send, while the packet poller monitors packet arrivals and notifies other modules upon completion.

\input{fig_tex/atlas-framework}

\noindent\textbf{Simulator Development:}
The thermal analyzer is developed upon HotSpot-7.0~\cite{han20222} to model 3D-Accelerator's transient thermal behavior. 
Following Fig.~\ref{fig:chip}, we construct a multi-layer stack containing one logic die and a configurable number of DRAM dies.
It performs simulation using a grid-based formulation, with the grid resolution set to 128 by default. 
Specifically,
at time step $t$, let $\mathbf{T}_t$ denote the temperature distribution, $\mathbf{P}_t$ the power input over $[t, t+\Delta t]$, and $\Delta t$ the simulation time step.
The analyzer then computes the temperature distribution at the next time step, $\mathbf{T}_{t+\Delta t}$.
The transient thermal evolution is governed by the discretized heat equation:
$\mathbf{C}\frac{\mathbf{T}_{t+\Delta t}-\mathbf{T}_t}{\Delta t} + \mathbf{G}\mathbf{T}_{t+\Delta t} = \mathbf{P}_t,$
where $\mathbf{C}$ and $\mathbf{G}$ denote the thermal capacitance matrix and the thermal conductance matrix, respectively.

For the 3D-Accelerator simulator, we integrate the timing parameters provided by our industry collaborators into Ramulator2~\cite{luo2023ramulator} to model 3D-DRAM performance.
We also extend its internal memory controller with a tile-level command scheduling mechanism tailored for LLM workloads to maximize bandwidth utilization.
Given that compute logic has diverse microarchitectures and organization granularities, while its execution latency is highly deterministic,
we use performance models based on FLOPs and SRAM traffic to
implement the logic-die simulator.
The inter-core interconnect simulator is built upon BookSim2~\cite{jiang2013detailed}, where we expose interfaces for both core simulator objects and the global manager.
The remaining simulator logic is implemented in \textasciitilde5.4K lines of C++ code.

\noindent\textbf{Extending \TitleAbbr:}
Although \TitleAbbr~currently targets LLM inference, a dominant application today, it supports 
to flexibly extend 
both architecture and operators, enabling broader applicability.
For architecture, we provide a unified \texttt{Component} base class to support consistent extension for all microarchitecture modules.
Replacing a component only requires connecting new implementation's cycle-level clock and input interfaces to corresponding base-class interfaces.
Adding a new component further requires updating architecture parameters and tiling explorer's  execution description generation based on the associated programming primitives.
For operators, our core-level execution semantics align with existing DSLs~\cite{wang2025tilelang,tillet2019triton}, natively supporting arbitrary intra-core execution. 
In addition, our MPMD model supports to express both communication and computation. 
By specifying \texttt{core\_id} parameter in kernel definition, arbitrary multi-core workloads can be
described.

\subsection{Simulator Fidelity Validation}
\label{sec:simulator-validation}

To validate the accuracy of the 3D-Accelerator simulator, we compare its reported performance with a 
3D-DRAM test chip, whose architecture parameters are summarized in Table~\ref{tab:architecture-config}.
We evaluate accuracy at three levels: (1) memory access performance in the 3D-DRAM simulator, (2) intra-core operator execution in the core simulator object, and (3) inter-core communication in the interconnect simulator.
We use production LLM workloads with realistic request scales.
Workload and operator implementation details are described in Sec.~\ref{sec:cloud-chip-dse}.
From these workloads, we collect all distinct memory-access traces and operator shapes, yielding 2304 DRAM access cases and 712 computation/inter-core communication cases.
The larger number of DRAM cases is because 3D-DRAM design exploration is conducted prior to 
chip design exploration in Sec.~\ref{sec:cloud-chip-dse}.
Besides, since each communication operator follows a compute operator, their case numbers are identical.

Fig.~\ref{fig:simulator-validation} shows the distribution of absolute errors.
For DRAM latency/bandwidth modeling, \TitleAbbr~achieves maximum errors of 7.11\%/7.65\%, and mean absolute error (MAE) of 3.83\%/4.01\%, with correlations of 99.61\%/98.49\% to measured performance.
For computation, the MAE is 2.16\%, with correlation 99.96\% and maximum error 8.21\%.
For inter-core communication, the MAE is 2.72\%, with correlation 97.26\% and maximum error 8.57\%.
Building on these results, for end-to-end inference, the absolute error is within 6.37\%.
These results prove the accuracy of the 3D-Accelerator simulator.

\input{table_tex/architecture-config}

\input{fig_tex/simulator-validation}

For thermal simulation, we inject thermophysical parameters measured on silico into the thermal analyzer to ensure accuracy.
We first prepare cross-sectional samples and use high-resolution scanning transmission electron microscopy (HR-STEM)~\cite{inkson2016scanning} to measure layer thickness, then expose each layer via mechanical polishing and measure thermal conductivity using the time-domain thermoreflectance (TDTR) test system in Table~\ref{tab:architecture-config} 
(component details listed in~\cite{jiang2018tutorial,huang2025first}).
The extracted parameters of different materials in 3D-Accelerator will be disclosed after publication.

%% file: fig_tex/3d-dram.tex
\begin{figure}
    \centering
    \includegraphics[width=0.99\columnwidth]{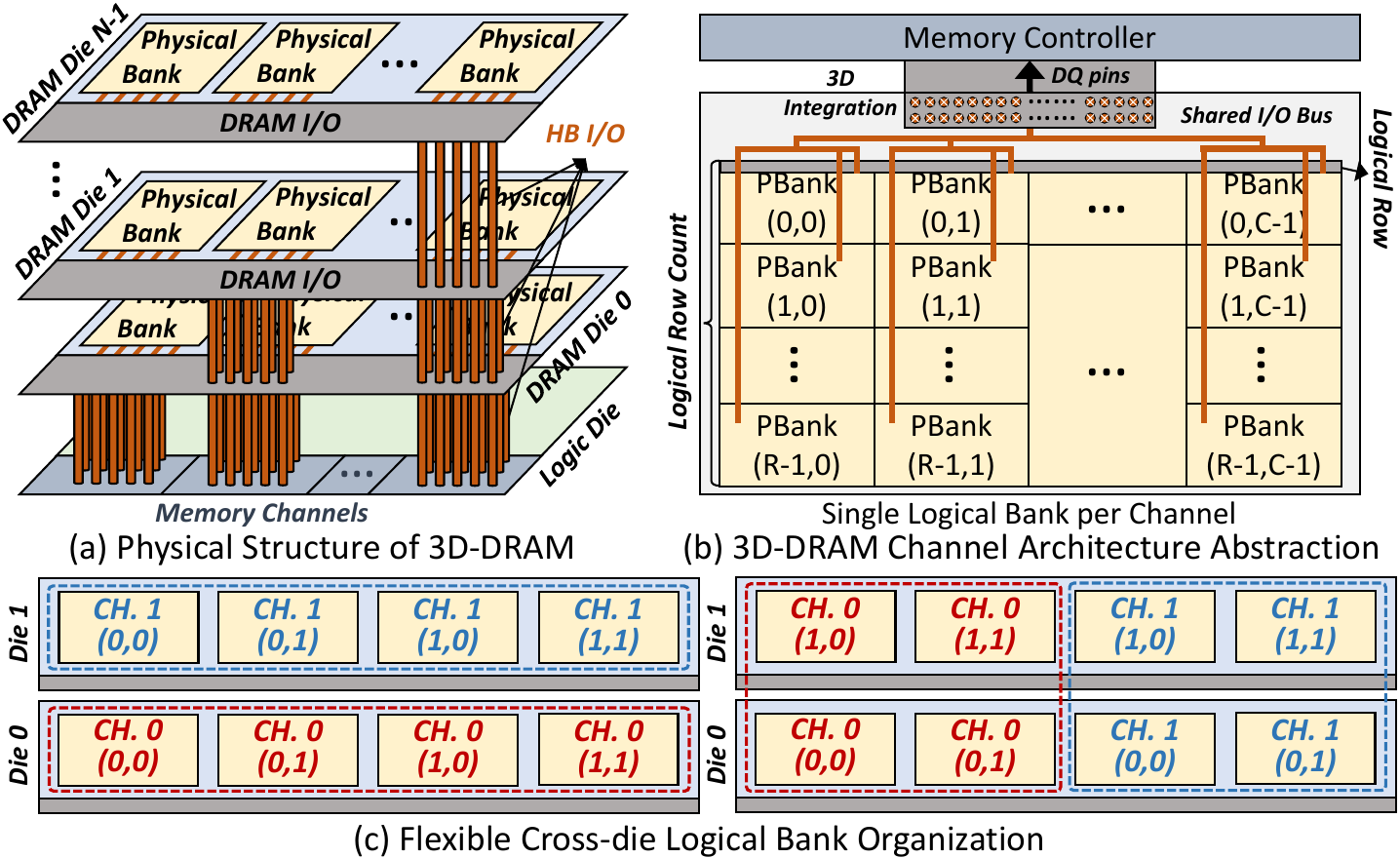}
    \caption{Architecture of Hybrid-Bonding-based 3D-DRAM.}
    \label{fig:3d-dram}
\end{figure}

%% file: fig_tex/chip.tex
\begin{figure}
    \centering
    \includegraphics[width=0.99\columnwidth]{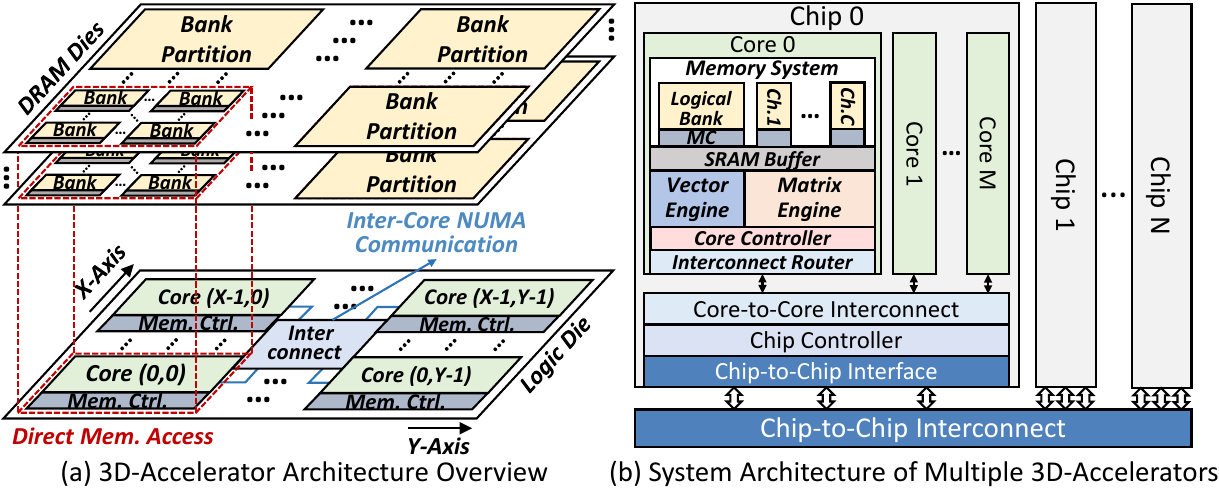}
    \caption{System Architecture of 3D-Accelerator.}
    \label{fig:chip}
\end{figure}

%% file: table_tex/architecture-template.tex
\begin{table}[t]
\centering
\caption{3D-Accelerator System Description Template.}
\label{tab:3d-accelerator-system}
\renewcommand{\arraystretch}{1}
\resizebox{0.475\textwidth}{!}{

\begin{tabular}{|c|c|c|c|c|}
\hline

\multicolumn{5}{|c|}{\textbf{3D-Accelerator System}} \\ \hline

\multirow{3}{*}{\begin{tabular}[c]{@{}c@{}}\textbf{Architecture}\\ \textbf{Hierarchy}\end{tabular}}
& \multicolumn{3}{c|}{\textbf{3D-Accelerator Architecture}}
& \multirow{3}{*}{\begin{tabular}[c]{@{}c@{}}\textbf{Inter-Accelerator}\\ \textbf{Interconnect}\end{tabular}} \\ \cline{2-4}

& \multicolumn{2}{c|}{\textbf{Core}}
& \multirow{2}{*}{\begin{tabular}[c]{@{}c@{}}\textbf{Inter-Core} \textbf{Interconnect}\end{tabular}}
&  \\ \cline{2-3}

& \textbf{3D-DRAM System}
& \textbf{Compute Logic}
&
& \\ \hline

\multirow{8}{*}{\begin{tabular}[c]{@{}c@{}}\textbf{Architecture}\\ \textbf{Specifications}\end{tabular}}

& \multirow{2}{*}{\begin{tabular}[c]{@{}c@{}}Physical Bank\\ Row Size (KB)\end{tabular}}
& \multirow{2}{*}{\begin{tabular}[c]{@{}c@{}}Matrix Compute\\ Capacity (TFLOPS)\end{tabular}}
& Core Array Size
& \multirow{4}{*}{Link Latency (s)} \\ \cline{4-4}

&
& 
& Topology 
& \\ \cline{2-4}

& Physical Bank Row Count
& \multirow{2}{*}{\begin{tabular}[c]{@{}c@{}}Vector Compute\\ Capacity (TFLOPS)\end{tabular}}
& \multirow{2}{*}{Link Latency (s)}
& \\ \cline{2-2}

& Logical Row Size ($C$)
&
&
& \\ \cline{2-5}

& Logical Row Count ($R$)
& \multirow{2}{*}{\begin{tabular}[c]{@{}c@{}}SRAM Buffer\\ Bandwidth (B/Cycle)\end{tabular}}
& \multirow{4}{*}{\begin{tabular}[c]{@{}c@{}}Interconnect\\ Bandwidth (GB/s)\\ (Link Width, Flit Size, etc.)\end{tabular}}
& \multirow{4}{*}{\begin{tabular}[c]{@{}c@{}}Interconnect Bandwidth (GB/s)\end{tabular}} \\ \cline{2-2}

& I/O pin Data Rate (Gbps)
&
&
& \\ \cline{2-3}

& I/O pin count per Channel
& SRAM Buffer Size (MB)
&
& \\ \cline{2-3}

& Channel Number
& Frequency (GHz)
&
& \\

\hline

\end{tabular}

}

\end{table}

%% file: table_tex/programming-interface.tex
\begin{table}[t]
\centering
\scriptsize
\renewcommand{\arraystretch}{1.}
\caption{Key Programming Primitives for 3D-Accelerators.}
\label{tab:programming-interface}
\resizebox{0.475\textwidth}{!}{%
\begin{tabular}{|c|c|c|}
\hline
\textbf{Primitive Scope} &
\textbf{Primitive Category} &
\textbf{Primitives} \\
\hline

\multirow{6}{*}{\makecell[c]{\textbf{Core-Level}\\\textbf{Execution}}}
& DRAM Tensor Declaration
& \texttt{\interface{tensor}(shape, stride, dtype)} \\
\cline{2-3}

& SRAM Tile Declaration
& \texttt{\interface{alloc}(shape, dtype)} \\
\cline{2-3}

& DRAM/SRAM Data Transfer
& \texttt{\interface{copy}(src, dst)} \\
\cline{2-3}

& Tensor Computation
& \texttt{\interface{gemm}(A, B)} \\
\cline{2-3}

& Vector Reduction
& \texttt{\interface{reduce\_max}}, \texttt{\interface{reduce\_sum}}, etc. \\
\cline{2-3}

& Vector Element-wise
& \texttt{\interface{add}}, \texttt{\interface{sub}}, \texttt{\interface{mul}}, \texttt{\interface{div}}, \texttt{\interface{exp}}, etc. \\
\hline

\multirow{4}{*}{\makecell[c]{\textbf{Core-Level}\\\textbf{Communication}}}
& Core Organization Declaration
& \texttt{\interface{core\_array}(shape)} \\
\cline{2-3}

& Inter-Core GEMM Partition
&\texttt{\interface{split\_gemm}(M, K, N, core\_dim\_mapping)} \\
\cline{2-3}

& Inter-Core Attention Partition
& \texttt{\interface{split\_attention}(token\_slot\_list)} \\
\cline{2-3}

& Inter-Core Data Transfer
& \texttt{\interface{send}(src, dst, data)}, \texttt{\interface{recv}(src, dst, data)} \\
\hline
\end{tabular}%
}

\end{table}

%% file: fig_tex/computation-programming.tex
\begin{figure}
    \centering
    \includegraphics[width=0.99\columnwidth]{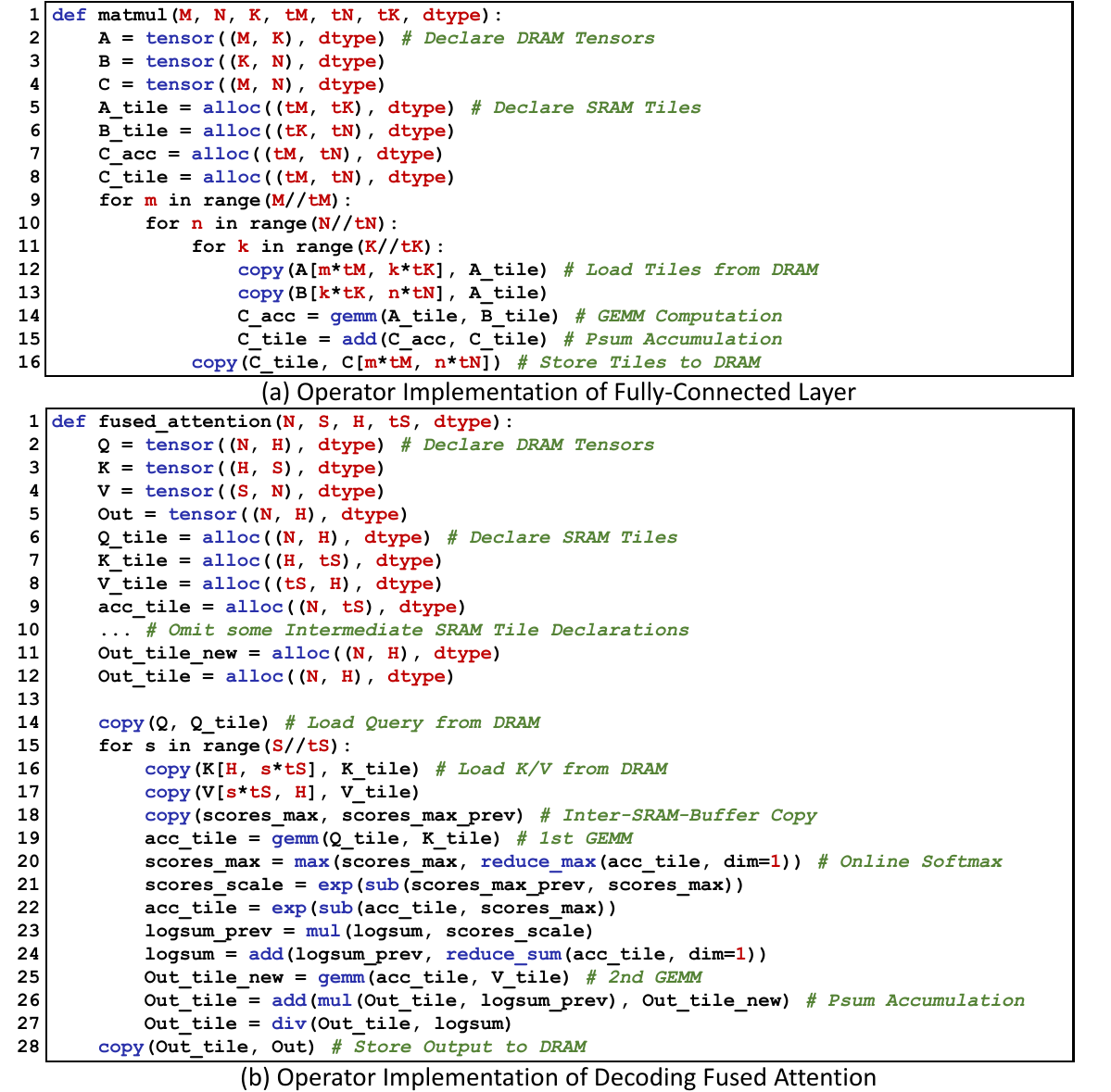}
    \caption{Core-Level Execution Programming.}
    \label{fig:computation-programming}
\end{figure}

%% file: fig_tex/communication-programming.tex
\begin{figure}
    \centering
    \includegraphics[width=0.99\columnwidth]{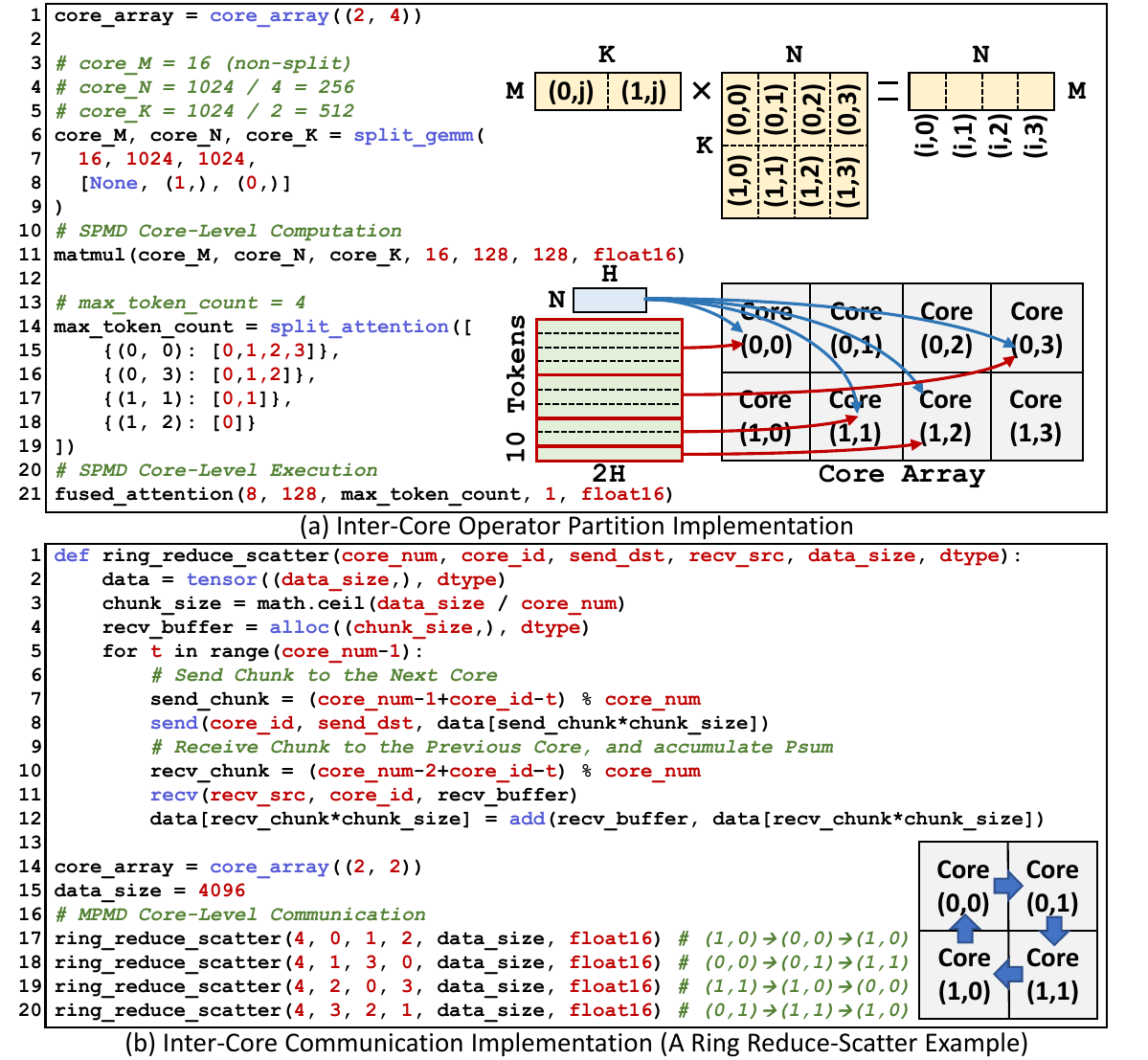}
    \caption{Core-Level Communication Programming.}
    \label{fig:communication-programming}
\end{figure}

%% file: fig_tex/atlas-framework.tex
\begin{figure}
    \centering
    \includegraphics[width=0.99\columnwidth]{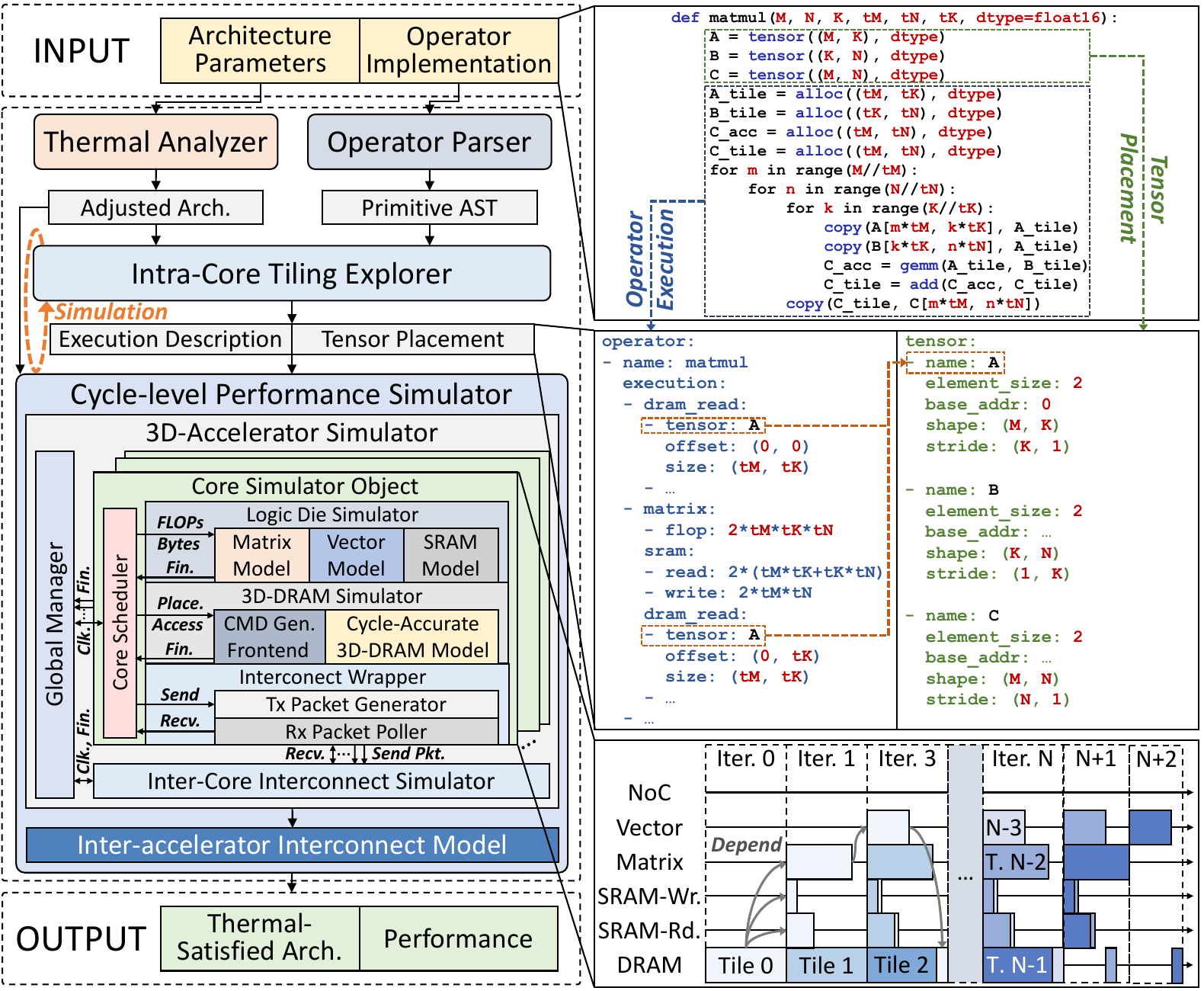}
    \caption{Overview of ATLAS Framework.}
    \label{fig:atlas-framework}
\end{figure}

%% file: table_tex/architecture-config.tex
\begin{table}[t]
\centering
\caption{Architecture Parameters of 3D-Accelerator Chip}
\label{tab:architecture-config}

\resizebox{0.475\textwidth}{!}{
\begin{tabular}{|c|c|c|}
\hline

\multirow{4}{*}{\begin{tabular}[c]{@{}c@{}}\textbf{3D-DRAM}\\ \textbf{Stack}\end{tabular}}

& Area 
& 800mm$^2$ $\times$ 4 Dies, 8192 physical banks per DRAM die \\ \cline{2-3}

& Physical Bank 
& \begin{tabular}[c]{@{}c@{}}2.5MB (2KB/row $\times$ 1280 rows)\end{tabular} \\ \cline{2-3}

& Logical Bank 
& 320MB, 4 physical bank rows ($R$=4), 32 physical banks per row ($C$=32)  \\ \cline{2-3}

& Data Rate 
& \begin{tabular}[c]{@{}c@{}}1024 pins/channel, 0.5Gbps/pin (64GB/s per channel)\end{tabular} \\ 

\hline

\multirow{5}{*}{\begin{tabular}[c]{@{}c@{}}\textbf{Logic} \textbf{Die}\end{tabular}}

& Area 
& 800mm$^2$, 7nm technology, 16 cores in 4 $\times$ 4 array \\ \cline{2-3}

& Performance 
& \begin{tabular}[c]{@{}c@{}}1GHz, 253.44TFLOPS (FP16), 16TB/s 3D-DRAM bandwidth, 242.24W peak power\end{tabular} \\ \cline{2-3}

& Core 
& \begin{tabular}[c]{@{}c@{}} 45.43mm$^2$, 15.84TFLOPS (15.36TFLOPS tensor + 0.48TFLOPS vector)\\
4MB SRAM buffer, 16 3D-DRAM channels (5GB capacity, 1TB/s bandwidth)\end{tabular} \\ \cline{2-3}

& NoC 
& Mesh topology, link width = 128B \\

\hline

\multirow{1}{*}{\begin{tabular}[c]{@{}c@{}}\textbf{Equipment}\end{tabular}}

& \begin{tabular}[c]{@{}c@{}} TDTR\\
Test System\end{tabular} 
& \begin{minipage}{1.\linewidth}
    \centering
    \vspace{1pt}
    \raisebox{-2pt}{\includegraphics[width=0.95\linewidth]{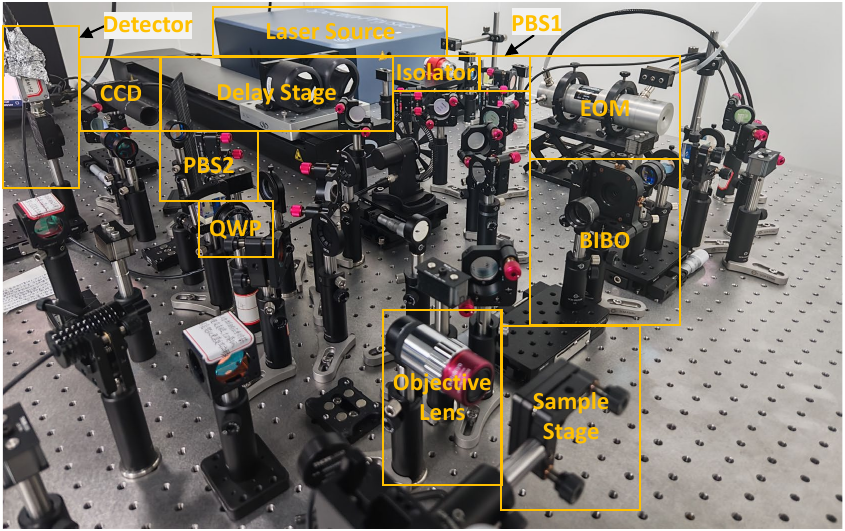}}
    \vspace{1pt}
\end{minipage} \\

\hline
\end{tabular}
}

\end{table}

%% file: fig_tex/simulator-validation.tex
\begin{figure}
    \centering
    \includegraphics[width=0.99\columnwidth]{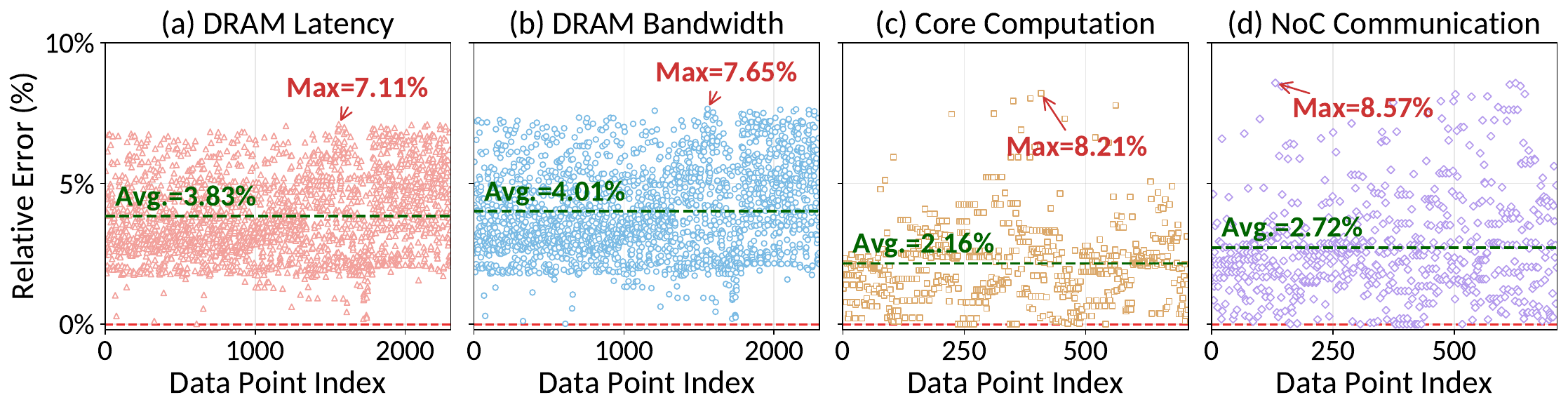}
    \caption{\TitleAbbr~Performance Simulator Fidelity Validation.}
    \label{fig:simulator-validation}
\end{figure}

%% file: tex/4-cloud-chip-dse.tex
\section{Optimizing Design for Cloud LLM Inference}
\label{sec:cloud-chip-dse}

After validating \TitleAbbr, we first optimize 3D-Accelerator for cloud inference to reveal how the parameters in Table~\ref{tab:architecture-config} are selected.
We begin with the design space exploration (DSE) of 3D-DRAM memory system, followed by the DSE for 3D-Accelerator architecture.

\subsection{Evaluation Setup}
\label{sec:cloud-evaluation-setup}

\input{fig_tex/dram-benchmark}

\noindent\textbf{Benchmarks:}
We construct DRAM benchmarks for 3D-DRAM DSE by extracting memory traces from core LLM operators (i.e., FC and attention).
For FC, matrices are stored contiguously (row/column-wise) but accessed tile by tile following the execution order. 
For example, in Fig.~\ref{fig:dram-benchmark}-(a), the matrix is laid out row-wise, while tiles of size $(tH_{in}, tH_{out})$ are accessed column-wise.
Since address continuity exists only within rows/columns inside a tile, DRAM organization determines row-buffer reuse and thus affects access performance.
For attention, practical cloud deployments manage the KV cache at block granularity to support dynamic request arrivals and evictions~\cite{kwon2023efficient,sglang}.
As shown in Fig.~\ref{fig:dram-benchmark}-(b), each core maintains $N$ blocks with $b$ slots per block ($b$=4 in Fig.~\ref{fig:dram-benchmark}-(b)). Each slot stores a token's KV vector. 
Data are laid out contiguously along the KV vector dimension, while the blocks accessed by a request can be non-contiguous due to request dynamicity (e.g., blocks $0$,$2$,$n_1$,$n_2$ in Fig.~\ref{fig:dram-benchmark}-(b)).

For 3D-Accelerator DSE, we use OPT-66B~\cite{zhang2022opt}, LLaMA3-70B~\cite{dubey2024llama}, Mixtral-8×22B~\cite{jiang2024mixtral}, and Qwen3-235B-A22B~\cite{yang2025qwen3} (all under FP16).
They cover dense (OPT/LLaMA) and MoE models (8-of-2 Mixtral, 128-of-8 Qwen), and span query-to-KV head ratios from 1:1 to 16:1.
As prefill-decoding disaggregation has become the standard deployment strategy in production~\cite{patel2024splitwise,qin2025mooncake,cai2026characterizing}, we focus on decoding evaluation to fully utilize 3D-Accelerator's high bandwidth.
We evaluate batch sizes (BS) of 16/64 to represent low/high-concurrency scenarios, and set context lengths to 1K/4K (OPT/Qwen) and 8K/32K (LLaMA/Mixtral) based on capacity requirements.

\input{fig_tex/llm-inference-dataflow}

Fig.~\ref{fig:decoding-dataflow} depicts 3D-Accelerator's decoding dataflow.
We map FC operators onto a 2D core array following the physical 2D-mesh layout.
Similar to Fig.~\ref{fig:communication-programming}-(a), each weight matrix dimension is split along one corresponding core-array dimension, while the batch dimension is unpartitioned to support arbitrary batch sizes and avoid weight replication.
For attention, to ensure load balance, we evenly split request contexts across all cores to perform fused attention.
For inter-core communication between consecutive FC operators, we follow prior work~\cite{pope2023efficiently} and use 1D all-reduce.
For attention, since each core requires the full query vector and produces partial sums, we use 2D all-reduce.
Inter-core communication is implemented with the TidalMesh algorithm~\cite{tidalmesh}, which outperforms ring-/tree-based schemes on 2D-mesh.
When models are partitioned across devices via tensor/expert parallelism (TP/EP), we insert inter-device all-reduce/all-to-all after attention blocks and FFN/MoE blocks.

\noindent\textbf{Baselines:}
After DSE, we compare our design with the following baselines:
(1) H200 GPU~\cite{h200}. Under the same area budget as our 3D-Accelerator (single-reticle for both compute and DRAM), it adopts 4nm logic process with HBM3E, providing 989TFLOPS FP16 compute and 4.8TB/s bandwidth, with 700W power.
(2) Stratum~\cite{pan2025stratum}.
As its monolithic DRAM has not yet been mass-produced, we emulate its configuration in our design space using the corresponding architecture parameters (details in Sec.~\ref{sec:cloud-arch-dse}). 
LLMs are deployed on an 8-device system with 900GB/s NVLink (empowered by NVLink Fusion~\cite{nvlink}), using TP=8 (and EP=8 for MoE models).

\input{fig_tex/cloud-dram-dse}

\subsection{3D-DRAM Architecture Exploration}
\label{sec:cloud-3d-dram-dse}

We first optimize per-core 3D-DRAM memory system under 
the hardware budgets in Table~\ref{tab:architecture-config} (2048 PBs + 16384 I/O pins per core),
ensuring constant core area, HB integration overhead, and peak DRAM bandwidth.
For workloads, since performance trends are consistent across GEMM shapes, we test (64,8192)×(8192,8192) ($M$=64, $K$=$N$=8192), using the partition in Sec.~\ref{sec:cloud-evaluation-setup}.
Memory traces follow the execution order in Fig.~\ref{fig:computation-programming}-(a), with $A$/$C$ stored in row-major and $B$ in column-major.
For attention, we vary block sizes and context lengths under KV vector length of 256 (used by all tested LLMs).
To stress-test the non-contiguous block layout caused by request dynamicity, we randomly generate block sequences and report results averaged over 10 runs.
We explore the following dimensions:

\noindent\textbf{Channel Interleaving:}
We first study how channel interleaving granularity affects bandwidth utilization under a given 3D-DRAM memory system.
As shown in Fig.~\ref{fig:dram-benchmark}-(c), to directly expose this effect,  we adopt a linear mapping scheme and vary consecutive byte count per row access ($2^x \cdot BL$).
Since similar bandwidth trends are observed across different memory systems, we evaluate a 16-channel design with 16KB logical rows (0$\le$$x$$\le$7).
As shown in Fig.~\ref{fig:cloud-dram-dse}-(a), bandwidth utilization increases as tile size (GEMM) or block size (attention) grows.
However, since tile/block sizes are chosen to balance bandwidth gains against non-overlapped prologue/epilogue overheads (Fig.~\ref{fig:atlas-framework}, Iter.0 and Iter.$N\!+\!1$/$N\!+\!2$) in practice, we should evaluate over the full performance distribution, rather than at extreme sizes.

In Fig.~\ref{fig:cloud-dram-dse}-(a), bandwidth utilization improves from $x$=0 to $x$=5 for both workloads, then saturates for GEMM and degrades for attention.
This is due to a fundamental trade-off: 
larger granularity improves row buffer locality by increasing intra-row access size,
but reduces inter-channel parallelism by concentrating accesses in fewer channels, thus lowering effective bandwidth.
Therefore, a moderate granularity (e.g., $x$=5) is preferred to balance these effects.

\noindent{\setlength{\fboxsep}{1pt}\fbox{%
    \parbox[t]{0.4675\textwidth}{%
        \strut
        \textbf{Takeaway 1}: Larger memory access granularity (tile/block size) improves 3D-DRAM bandwidth utilization.\\
        \textbf{Takeaway 2}: Channel interleaving granularity requires a balance: too small hurts
        row locality, too large reduces channel parallelism.
    }
}}%

\noindent\textbf{I/O Pin Organization:}
Then, we study the impact when grouping I/O pins into different channel counts (ch.).
For each ch., we fix logical row size to 16KB and report the best performance across interleaving settings.
In Fig.~\ref{fig:cloud-dram-dse}-(b), utilization improves from 2ch. to 16ch. across tile/block sizes for both GEMM (17.4-46.9 percentage points (pp)) and attention (13.8-43.7pp).
When ch.$\ge$16, the gains diminish: GEMM improves marginally at large tile sizes ($\le$8.2pp) but degrades sharply at moderate sizes (up to 39.1pp), while attention benefits only very small block sizes ($\le$4).
These trends arise from a trade-off:
More channels increase concurrency by distributing accesses (tile rows/columns, KV blocks) across channels.
However, under fixed interleaving, activating more channels requires larger access sizes.
Small and consecutive accesses (e.g., moderate GEMM tiles) fails to utilize all channels, reducing effective bandwidth.
Finer-grained interleaving mitigates this issue but still hurts performance as discussed above.
Therefore, excessive channels are counterproductive, and a moderate count (e.g., 16) strikes the best balance.

\noindent{\setlength{\fboxsep}{2.5pt}\fbox{%
    \parbox[t]{0.46\textwidth}{%
        \textbf{Takeaway 3}: Increasing channels improves concurrency, but excessive channels reduce
        channel utilization and hurt bandwidth.
    }
}}%

\noindent\textbf{Logical Row Size:}
We next study the impact of logical row size under 16 channels (128 PBs per channel), and report the best performance across interleaving settings.
As shown in Fig.~\ref{fig:cloud-dram-dse}-(c), bandwidth utilization for both GEMM and attention increases as the logical row size grows, with gains up to 60.7pp and 37.7pp, respectively.
This is because larger rows increase the chance that accesses hit the same row, improving row buffer locality and reduces inter-row ACT/PRE operations.
Accordingly, increasing the logical row size (e.g., 32-64KB) is beneficial for 3D-DRAM performance.

\noindent{\setlength{\fboxsep}{1pt}\fbox{%
    \parbox[t]{0.4675\textwidth}{%
        \textbf{Takeaway 4}: Larger logical row sizes improve bandwidth utilization by boosting row buffer 
        locality and reducing ACT/PRE.
    }
}}%

\noindent\textbf{End-to-End Evaluation:}
Finally, we evaluate all design points under end-to-end LLM decoding.
As trends are consistent across batch sizes and context lengths, we report results for BS=64 and context length 4K/16K, along with latency-weighted average across all LLMs.
For each (channel count, logical row size), we report the best performance over all interleaving and tiling strategies.
As shown in Fig.~\ref{fig:cloud-dram-dse}-(d), LLaMA/Mixtral/Qwen achieve near-optimal performance with 8/16ch., while OPT benefits from more channels due to its large FFN hidden dim (36864). 
Increasing logical row size consistently improves utilization, and 64KB rows with 8/16ch. perform best on average.
Since 16ch. performs better on more models (OPT/LLaMA/Qwen), we select this design with its optimal interleaving ($x$=5), incurring 4.99mm$^2$ logic die area overhead.

\input{fig_tex/cloud-comp-bw-dse}

\subsection{3D-Accelerator Architecture Exploration}
\label{sec:cloud-arch-dse}

After determining the 3D-DRAM design, we next explore the 3D-Accelerator architecture.
All hardware designs are synthesized under the same 7nm process as Table~\ref{tab:architecture-config}, with identical logic die area budgets.
To control inter-core communication overhead and ensure feasible P\&R for distributed 3D-DRAM controllers, we fix the core array size to 4×4.
NoC topology is 2D-mesh due to its maturity and wide adoption in industry products~\cite{talpes2022dojo,abts2022groq,prabhakar2024sambanova,lie2022cerebras}. 
For thermal management, we adopt a copper liquid cooling plate with heat transfer coefficient (HTC) of 10000W/(m$^2\cdot$K)~\cite{yu2025cramming} and constrain chip temperature below 85\textcelsius{} for retention~\cite{yue20253d,yue2024exploiting}. 
If the thermal limit is exceeded, we reduce frequency (default 1GHz, down to 0.1GHz) or scale down compute area. 
Under these settings, we explore the following design dimensions 
through a hierarchical control strategy:
bandwidth allocation is evaluated under the best settings of all other dimensions.
Subsequent dimensions (SRAM, matrix-vector ratio, NoC link width) progressively fix prior choices, with SRAM and matrix-vector studies additionally fixing the link width to 128B.

\noindent\textbf{Bandwidth Allocation:}
We first explore bandwidth allocation by varying per-core channel count (ch.) under the above channel organization
(1024 pins, 64KB logical rows).
To keep capacity and DRAM timing constant, we decrease PB row count ($R$) as ch. increases, and vice versa.
Increasing ch. scales bandwidth proportionally, but enlarges MC area and reduces compute area~\cite{li2025h2}.
Moreover, under a fixed thermal budget, higher DRAM power reduces compute power budget, further limiting compute capacity.
Fig.~\ref{fig:cloud-comp-bw-dse}-(a) shows the compute-bandwidth trade-off.
From 16ch. to 4ch., bandwidth is lowered by 75\%, but compute is raised by only 7.3\%.
However, 32/64ch. sharply reduce compute due to larger MC area and tighter thermal constraints.
To show the thermal impact, Fig.~\ref{fig:cloud-comp-bw-dse}-(b) reports peak temperature and the ratio between achieved and peak compute (using all available area at 1GHz).
At 32ch., reducing compute by 30\% barely meets the 85\textcelsius{} limit (84.9\textcelsius{}). 
At 64ch., even with the minimum compute-bandwidth ratio 
at 1GHz (1:1, 4TFLOPS), further reducing frequency to 0.1GHz still 
leads to 110.2\textcelsius{} temperature.

Fig.~\ref{fig:cloud-comp-bw-dse}-(d) shows decoding 
latency comparison results.
For dense models, 16ch. achieves the best performance: fewer ch. limits compute utilization, while more ch. reduces compute and hurts FC performance (despite benefiting attention).
For MoE models, 32ch. performs better due to smaller per-expert batch sizes after expert routing. 
However, as batch size increases, 16ch. achieves comparable performance.
To ensure robust performance across scenarios and reduce thermal pressure, we select the 16ch. design, achieving the best average speedup (2.53×) across all models (Fig.~\ref{fig:cloud-comp-bw-dse}-(c)).

\noindent{\setlength{\fboxsep}{1pt}\fbox{%
    \parbox[t]{0.4675\textwidth}{%
        \textbf{Takeaway 5}: Low bandwidth limits compute utilization, while excessive bandwidth reduces compute and stresses thermal limits.
    }
}}%

\input{fig_tex/cloud-sram-dse}

\noindent\textbf{SRAM Allocation:}
Next, we vary per-core SRAM size.
As shown in Fig.~\ref{fig:cloud-sram-dse}-(a), reducing SRAM from 4MB to 1MB lowers area by only 10.9\%, yielding limited compute gain (6.0\%).
This is because SRAM must provide sufficient bandwidth for compute.
As SRAM shrinks while I/O demand increases (due to higher compute capacity), more I/O must be provided by smaller SRAM banks, reducing area efficiency and leading to marginal area savings.
In contrast, increasing SRAM beyond 4MB significantly enlarges area and reduces compute capacity.
From comparisons in Fig.~\ref{fig:cloud-sram-dse}-(b)/(c), 4MB already provides sufficient data reuse for all models.
For Mixtral, with a huge per-chip hidden dimension after EP, moderately larger SRAM improves bandwidth utilization via bulk DRAM accesses, yielding up to 7.2\% speedup.
However, beyond 4MB, performance drops as the reduced compute outweighs the memory benefit.
We therefore select 4MB per core to balance area efficiency and bandwidth benefits.

\noindent{\setlength{\fboxsep}{1pt}\fbox{%
    \parbox[t]{0.4675\textwidth}{%
        \textbf{Takeaway 6}: Moderate SRAM size balances area efficiency under on-chip bandwidth demand and compute capacity provision.
    }
}}%

\input{fig_tex/cloud-matrix-vector-dse}

\input{table_tex/stratum-config}

\noindent\textbf{Matrix-Vector Compute Allocation:}
As shown in Fig.~\ref{fig:computation-programming}-(b), fused attention introduces a bulk of vector operations in online softmax.
Therefore, we explore the matrix-vector compute ratio, providing different compute capacity settings in Fig.~\ref{fig:cloud-matrix-vector-dse}-(a).
From the latency comparisons in Fig.~\ref{fig:cloud-matrix-vector-dse}-(c), 16:1/32:1 achieve the highest speedup and outperform extreme settings (64:1/4:1) by up to 1.11/1.24×, respectively.
To further illustrate operator-level effects, Fig.~\ref{fig:cloud-matrix-vector-dse}-(c) (bottom) breaks down the speedup of FC (bars) and attention (lines), each normalized to its slowest configuration.
FC benefits from increasing matrix compute, but the gain slows under area constraints and saturates around 32:1.
In contrast, attention requires stronger vector compute and is effectively accelerated within 8:1-32:1.
To balance the gains across operators, we select 32:1 as the final design, achieving the highest average speedup as shown in Fig.~\ref{fig:cloud-matrix-vector-dse}-(b).

\noindent{\setlength{\fboxsep}{1pt}\fbox{%
    \parbox[t]{0.4675\textwidth}{%
        \textbf{Takeaway 7}: Balanced matrix-vector compute allocation matches both matrix-heavy FC and vector-heavy attention.}
}}%

\noindent\textbf{NoC Resource Allocation:}
Finally, we vary link width on the 4×4 2D-mesh.
Fig.~\ref{fig:cloud-noc-dse}-(a) shows the trade-off between link width and compute capacity under the fixed area budget.
From the evaluation in Fig.~\ref{fig:cloud-noc-dse}-(c) (top), link widths of 128/256B achieve the best speedup.
To further analyze performance, Fig.~\ref{fig:cloud-noc-dse}-(c) (bottom) shows communication speedup (bars) and compute share in decoding latency (lines).
Communication is steadily accelerated with wider links, while its latency share drops to \textasciitilde20\% once link width $\ge$ 128B, limiting further end-to-end gains.
We therefore select 128B link width to achieve the best end-to-end speedup as shown in Fig.~\ref{fig:cloud-noc-dse}-(b).

\noindent{\setlength{\fboxsep}{1pt}\fbox{%
    \parbox[t]{0.4675\textwidth}{%
        \textbf{Takeaway 8}: Moderate link width balances the efficiency in intra-core computation and inter-core communication.}
}}%

\input{fig_tex/cloud-noc-dse}

\input{fig_tex/cloud-baseline-comparison}

\noindent\textbf{Comparison vs. Baselines:}
To show the gains of our DSE-derived design, we compare it with H200 and Stratum.
For fair comparison, Stratum is instantiated under the same settings as our DSE, using its per-core parameters (\textasciitilde2TB/s bandwidth, 2.25MB SRAM, 32:1 matrix-vector, 128B link width).
During evaluation, we reserve another 128B NoC (2D-mesh) and 4MB SRAM in each core to decouple inter-/intra-accelerator communication.
As listed in Table~\ref{tab:stratum-config}, Stratum's higher bandwidth increases power and pushes temperature to 85\textcelsius{}.
To meet this constraint, we tune compute area and frequency to maximize achievable compute.
The best design runs at 1GHz but leaves 2.67mm$^2$ unused area, achieving only 59.2\% of theoretical peak compute.
Besides, its 2.25MB buffer saves only \textasciitilde1mm$^2$ area due to bandwidth demand.
Despite using the same matrix-vector ratio and NoC link width, Stratum’s excessive bandwidth and smaller SRAM hurt area efficiency and worsen thermal behavior.

Fig.\ref{fig:cloud-baseline-comparison} compares decoding speedup and energy efficiency.
Similar to bandwidth allocation DSE, \TitleAbbr~outperforms Stratum on dense models (up to 1.42×)
and achieves comparable performance on MoE at BS=64 (0.88-1.27×), 
while Stratum performs better on MoE at BS=16 (up to 1.39×).
For energy efficiency, \TitleAbbr~outperforms Stratum on dense models and MoE at BS=64 (up to 1.91×), and and matches it on MoE at BS=16 (0.93-0.97×).
Overall, \TitleAbbr~achieves 2.53/1.08× average speedup and 6.66/1.73× average energy efficiency over H200/Stratum, respectively.

%% file: fig_tex/dram-benchmark.tex
\begin{figure}
    \centering
    \includegraphics[width=0.99\columnwidth]{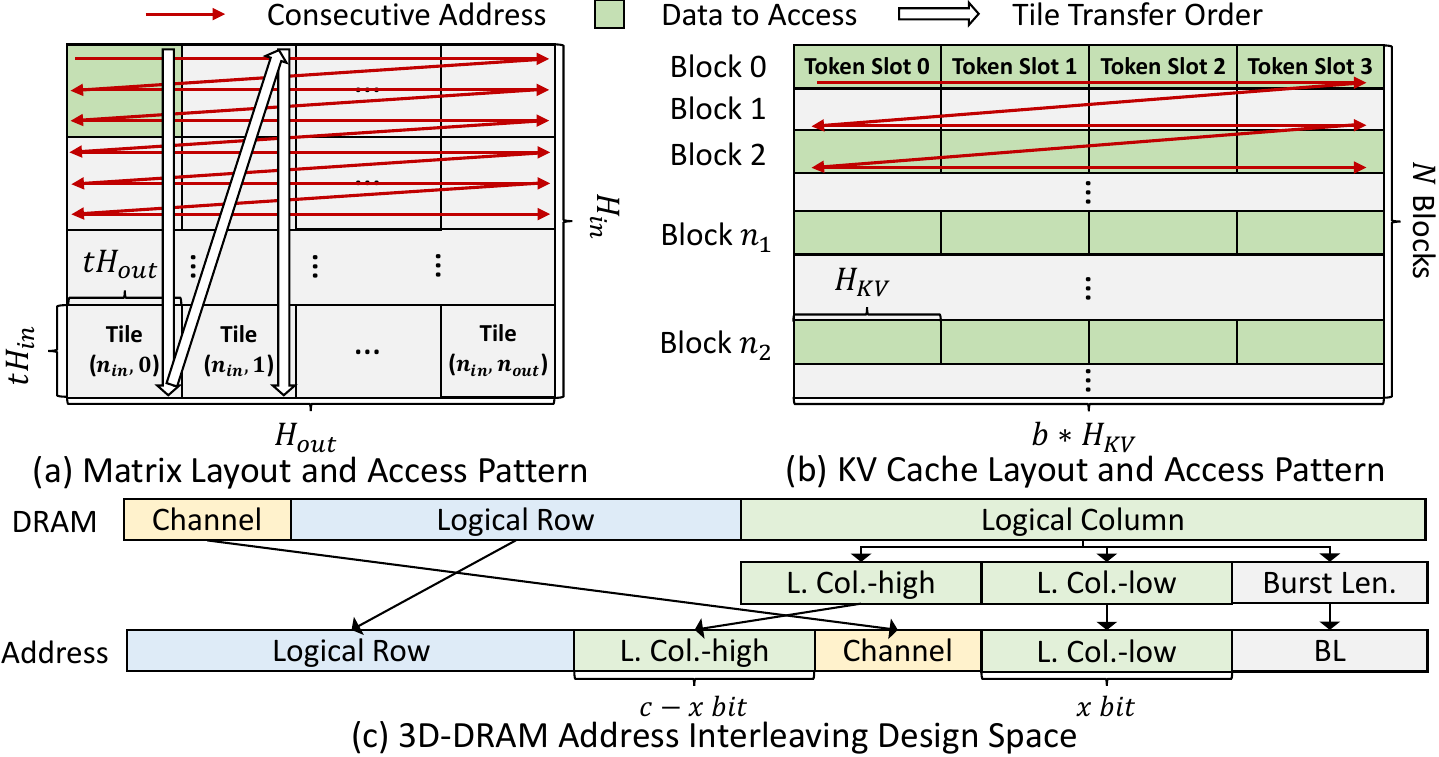}
    \caption{3D-DRAM Benchmark and Channel Interleaving.}
     \vspace{-0.5em}
    \label{fig:dram-benchmark}
\end{figure}

%% file: fig_tex/llm-inference-dataflow.tex
\begin{figure}
    \centering
    \includegraphics[width=0.99\columnwidth]{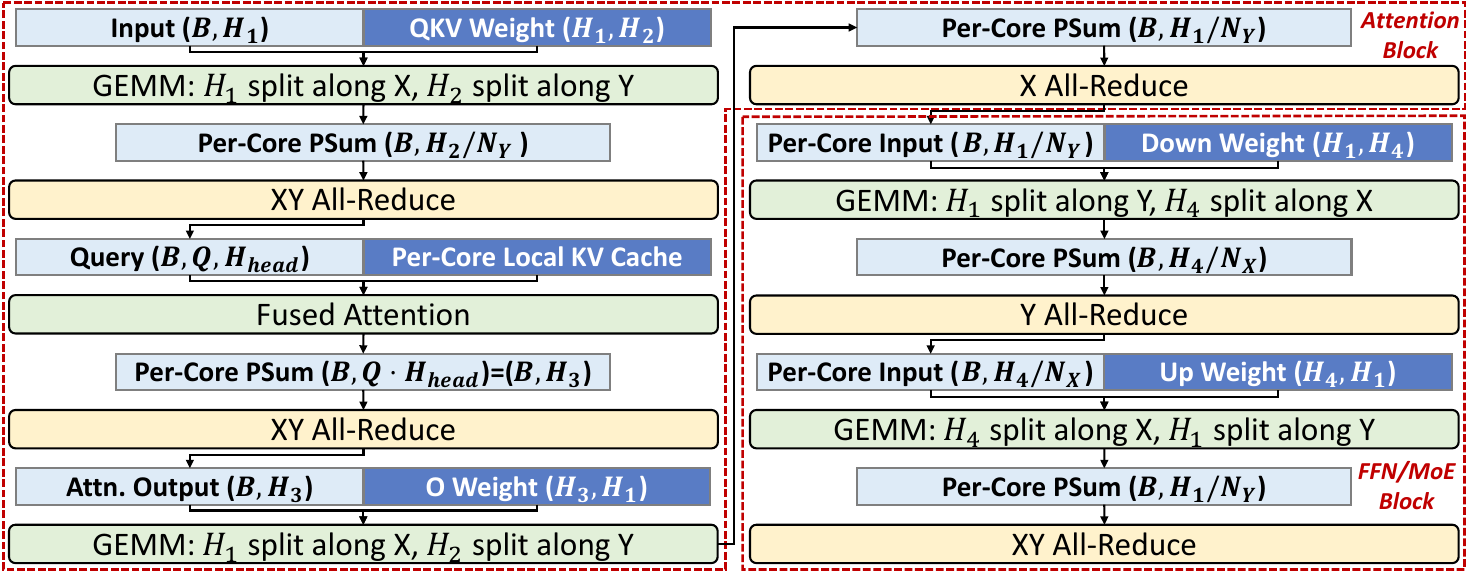}
    \caption{LLM Decoding Dataflow on Cloud 3D-Accelerator.}
    \label{fig:decoding-dataflow}
    \vspace{-0.5em}
\end{figure}

%% file: fig_tex/cloud-dram-dse.tex
\begin{figure}
    \centering
    \includegraphics[width=0.99\columnwidth]{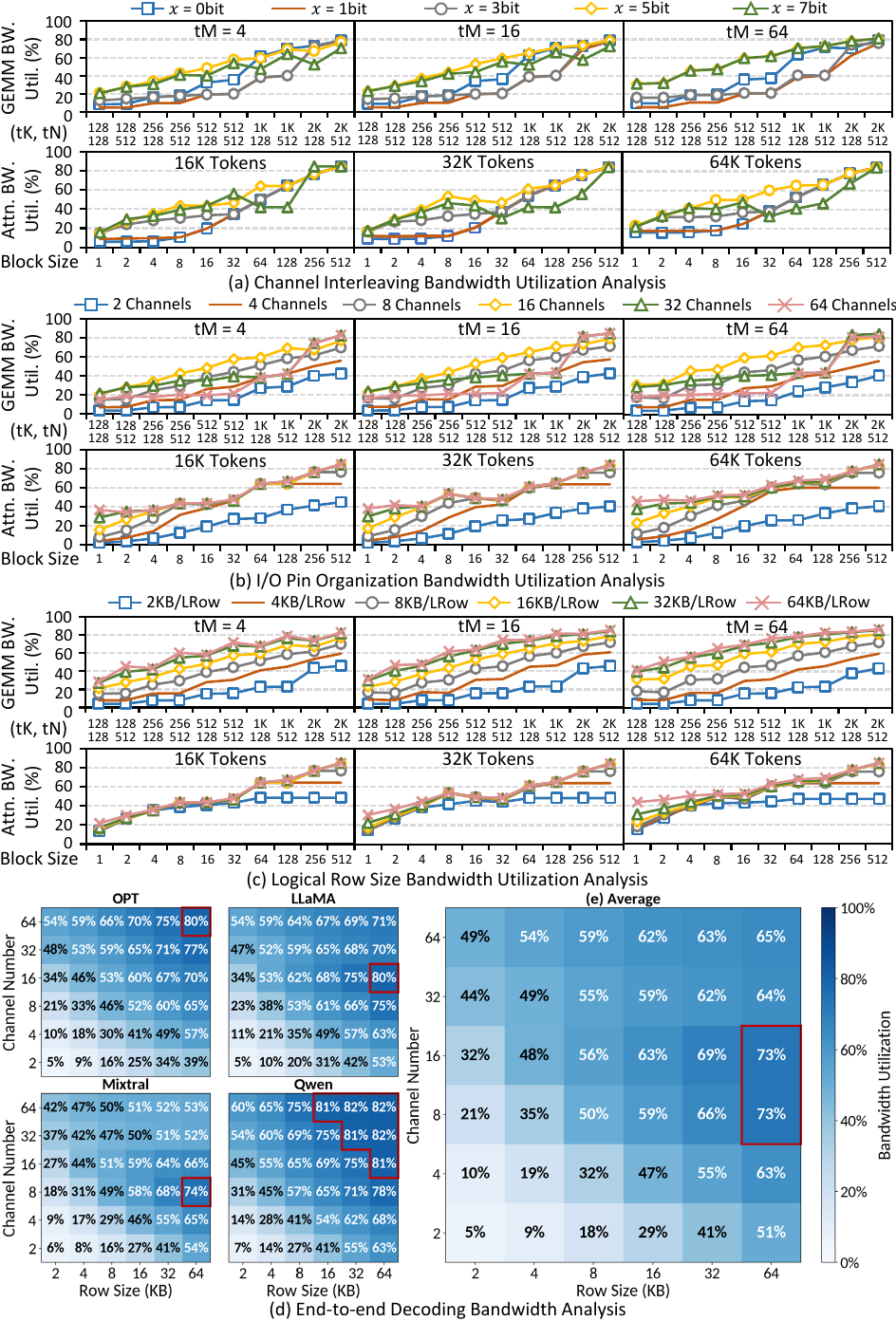}
    \caption{3D-DRAM Organization DSE for Cloud LLM.}
     \vspace{-0.5em}
    \label{fig:cloud-dram-dse}
\end{figure}

%% file: fig_tex/cloud-comp-bw-dse.tex
\begin{figure}
    \centering
    \includegraphics[width=0.99\columnwidth]{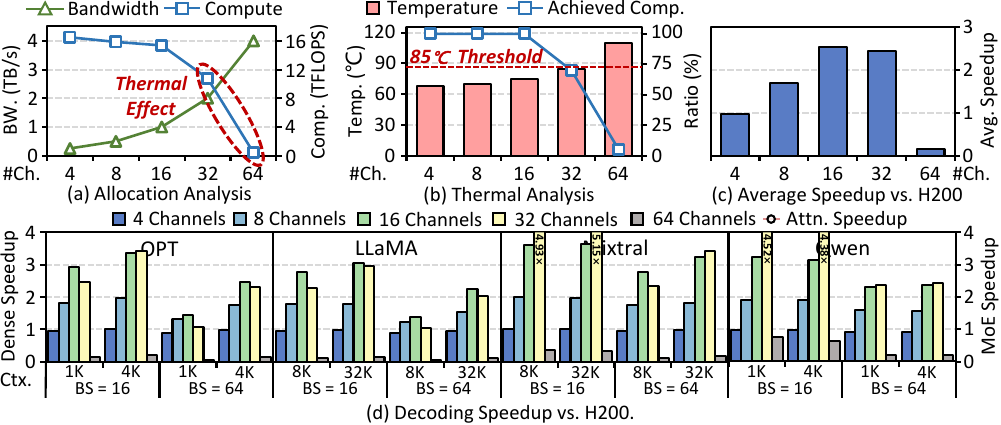}
    \caption{Bandwidth Allocation Analysis for Cloud LLM.}
    \label{fig:cloud-comp-bw-dse}
\end{figure}

%% file: fig_tex/cloud-sram-dse.tex
\begin{figure}
    \centering
    \includegraphics[width=0.99\columnwidth]{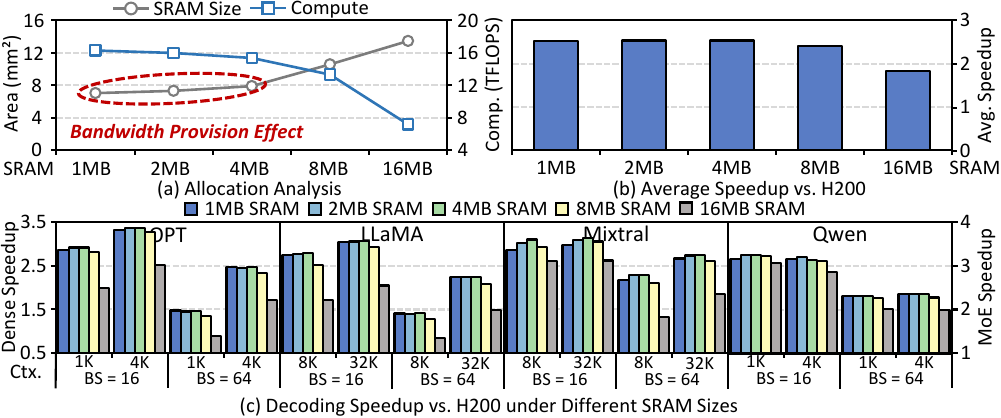}
    \caption{SRAM Buffer Size Analysis for Cloud LLM.}
    \label{fig:cloud-sram-dse}
\end{figure}

%% file: fig_tex/cloud-matrix-vector-dse.tex
\begin{figure}
    \centering
    \includegraphics[width=0.99\columnwidth]{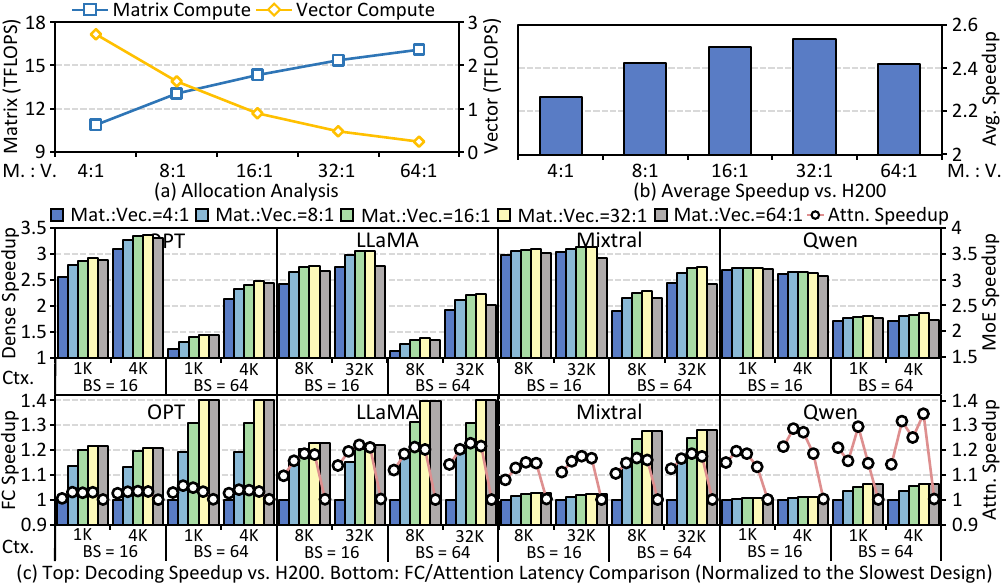}
    \caption{Matrix-Vector Allocation Analysis for Cloud LLM.}
    \label{fig:cloud-matrix-vector-dse}
\end{figure}

%% file: table_tex/stratum-config.tex
\begin{table}[t]
\centering
\caption{Architecture Parameters of Stratum-based Design}
\label{tab:stratum-config}

\resizebox{0.475\textwidth}{!}{
\begin{tabular}{|c|c|c|}
\hline

\multirow{4}{*}{\begin{tabular}[c]{@{}c@{}}\textbf{Stratum} \textbf{Logic Die}\end{tabular}}

& Performance 
& \begin{tabular}[c]{@{}c@{}}1GHz, 135.17TFLOPS (FP16), 32TB/s 3D-DRAM bandwidth, 325.77W peak power\end{tabular} \\ \cline{2-3}

& Core 
& \begin{tabular}[c]{@{}c@{}} 45.43mm$^2$, 8.45TFLOPS (8.19TFLOPS tensor + 0.26TFLOPS vector)\\
2.25MB SRAM buffer, 32 3D-DRAM channels (5GB capacity, 2TB/s bandwidth)\end{tabular} \\ \cline{2-3}

& NoC 
& Mesh topology, link width = 128B \\

\hline

\multirow{8.5}{*}{\begin{tabular}[c]{@{}c@{}}\textbf{Area, Power,}\\ \textbf{Thermal Analysis}\end{tabular}}

& \TitleAbbr
& \begin{minipage}{1.25\linewidth}
    \vspace{1pt}
    \centering
    \raisebox{-2pt}{\includegraphics[width=0.95\linewidth]{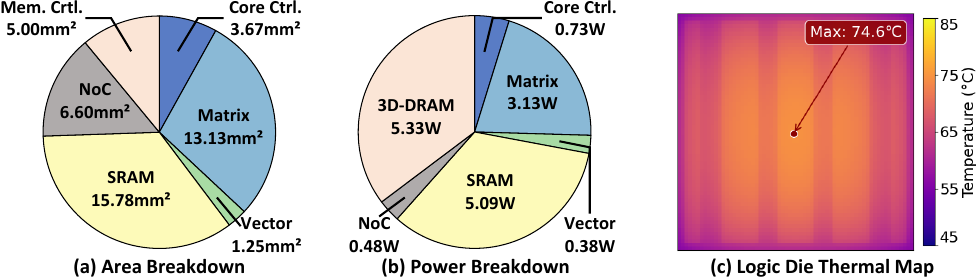}}
    \vspace{1pt}
\end{minipage}\\ \cline{2-3}

& Stratum
& \begin{minipage}{1.25\linewidth}
    \centering
    \vspace{1pt}
    \raisebox{-2pt}{\includegraphics[width=0.95\linewidth]{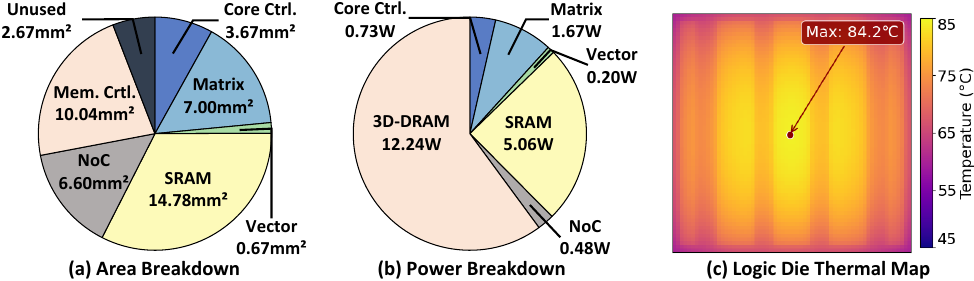}}
    \vspace{1pt}
\end{minipage}\\

\hline
\end{tabular}
}

\end{table}

%% file: fig_tex/cloud-noc-dse.tex
\begin{figure}
    \centering
    \includegraphics[width=0.99\columnwidth]{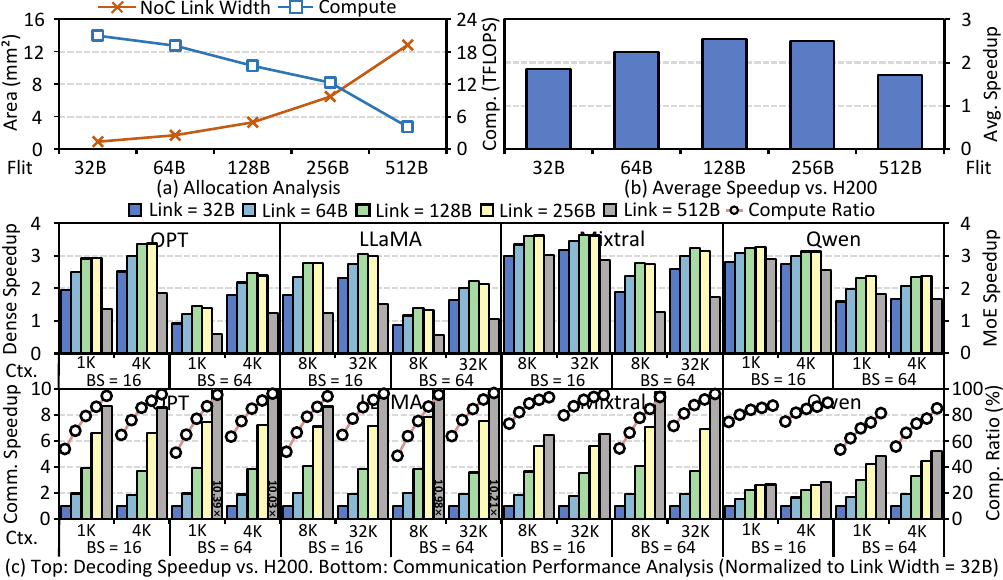}
    \caption{NoC Link Width Analysis for Cloud LLM.}
    \label{fig:cloud-noc-dse}
\end{figure}

%% file: fig_tex/cloud-baseline-comparison.tex
\begin{figure}
    \centering
    \includegraphics[width=0.99\columnwidth]{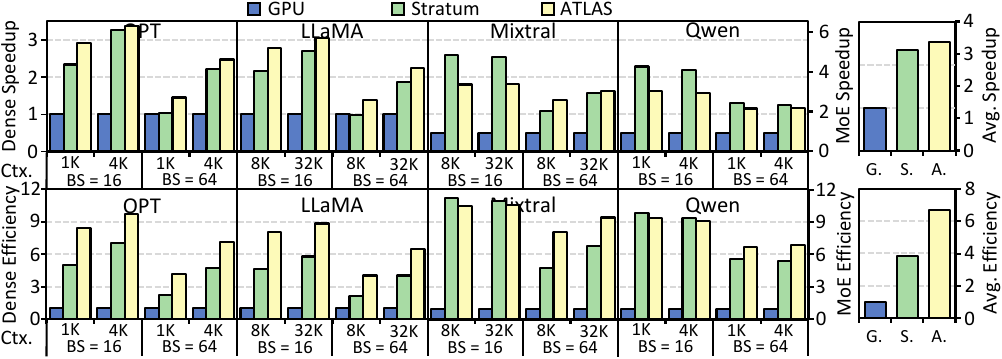}
    \caption{Latency and Energy Comparison vs. Baselines.}
    \label{fig:cloud-baseline-comparison}
\end{figure}

%% file: tex/5-edge-chip-dse.tex
\section{Optimizing Design for Edge LLM Inference}

In this section, we extend H$^2$-LLM~\cite{li2025h2} to optimize edge-side 3D-Accelerators, using TSMC-28nm logic die stacked with one DRAM die.
We adopt the same NPU configuration as H$^2$-LLM and follow its heterogeneous design with eight 3D-Accelerators, each connected via 12.8 GB/s LPDDR5 interfaces.
Each 3D-Accelerator contains 16 cores, with a per-core area of 6.76 mm$^2$.
We evaluate OPT/LLaMA3/PaLM-8B~\cite{zhang2022opt,dubey2024llama,chowdhery2022palm} with BS=1-16 and context lengths of 1K/2K.
To directly compare hardware designs, operators are executed sequentially, each following H$^2$-LLM's Operator Execution Flow and running across all 3D-Accelerators.
For thermal management, we use a heat spreader with HTC of 500W/(m$^2 \cdot $K)~\cite{kheirabadi2016cooling} and maintain the 85\textcelsius{} temperature constraint. 
Chip thickness and thermal conductivity are identical to the cloud 3D-Accelerator.

\input{fig_tex/edge-dram-dse}

\subsection{3D-DRAM Architecture Exploration}

Since H$^2$-LLM has thoroughly explored 3D-DRAM bandwidth provision but not the internal architecture, we study the design space in Sec.~\ref{sec:cloud-3d-dram-dse} under the same bandwidth settings as H$^2$-LLM (512pins at 0.4Gbps per core).
To match H$^2$-LLM's capacity (256MB/core), each core is equipped with 64 PBs of 4 MB each (2KB/row, 2K rows).
As performance trends are consistent across operator shapes, we evaluate LLaMA3-8B's FC (weight shape (4096,4096)) and GQA under BS=8 and context length 2K, following  H$^2$-LLM's partition strategy.
For FC, we vary hidden-dimension tiling within each core after H$^2$-LLM's inter-core partition, without tiling batch size due to its small value in edge inference.
Since H$^2$-LLM treats attention as two GEMMs, attention’s memory access latency is evaluated as the sum of the two GEMMs across tiling combinations. 

Fig.~\ref{fig:edge-dram-dse}-(a)\textasciitilde(c) shows 3D-DRAM DSE results.
We set 4ch. and 4KB logical rows by default, using optimal interleaving for I/O pin organization and logical row size DSE.
Unlike cloud designs, fine-grained interleaving ($x\le3$) is optimal, as smaller edge operators benefit less from long contiguous accesses.
For I/O pin organization, increasing channel count consistently improves performance.
This is because more channels imply fewer pins per channel, enabling finer-grained access and better matching small-scale edge workloads.
For logical row size, similar to cloud designs, increasing logical row size consistently improves performance.
Fig.~\ref{fig:edge-dram-dse}-(d) presents decoding bandwidth comparison 
\noindent{at} BS=4 and context length 2K, 
aligning with the trends as discussed above.
Accordingly, we select 8ch. and 16KB logical rows, with optimal interleaving $x$=1.
This design introduces 1.33mm$^2$ logic die area overhead under 28nm.

\noindent{\setlength{\fboxsep}{0.9pt}\fbox{%
    \parbox[t]{0.4675\textwidth}{%
        \textbf{Takeaway 9}: Fine-grained interleaving and more channels improve bandwidth utilization for small-scale edge workloads.}
}}%

\input{fig_tex/edge-matrix-vector-dse}

\subsection{3D-Accelerator Architecture Exploration}

We focus on matrix-vector allocation for edge 3D-Accelerators.
This is because H$^2$-LLM has thoroughly explored bandwidth and SRAM allocation, and edge workloads do not require NoC for complex inter-core communication.
H$^2$-LLM lacks vector units and avoids partitioning the input hidden dimension in each 3D-Accelerator to reduce partial-sum traffic through the external memory interface.
This constraint limits compute efficiency.
To expand operator tiling space, we introduce vector units to enable intra-accelerator partial-sum accumulation and reduce memory interface data movement.

Fig.~\ref{fig:edge-matrix-vector-dse}-(a) shows per-core compute distribution under the same SRAM setup as H$^2$-LLM (32KB input global buffer, per-core 32KB weight + 4KB output buffers).
From Fig.~\ref{fig:edge-matrix-vector-dse}-(c), under memory-bound workloads (BS=1/4), all designs benefit from the expanded tiling space (up to 1.09× speedup, 1.04× energy efficiency).
Under compute-bound workloads (BS=16), excessive vector provision hurts the performance, while moderate ratios (8:1-16:1) still outperforms H$^2$-LLM (up to 1.11× speedup, 1.17× energy efficiency).
We therefore select 8:1-ratio, achieving the best average performance (1.08× speedup, 1.09× energy efficiency in Fig.~\ref{fig:edge-matrix-vector-dse}-(b)).
All designs operate within 57.2-58.7\textcelsius{}, satisfying thermal constraints.

\noindent{\setlength{\fboxsep}{0.9pt}\fbox{%
    \parbox[t]{0.4675\textwidth}{%
        \textbf{Takeaway 10}: Moderate vector compute expands operator tiling flexibility and improves compute efficiency for edge workloads.}
}}%

%% file: fig_tex/edge-dram-dse.tex
\begin{figure}
    \centering
    \includegraphics[width=0.99\columnwidth]{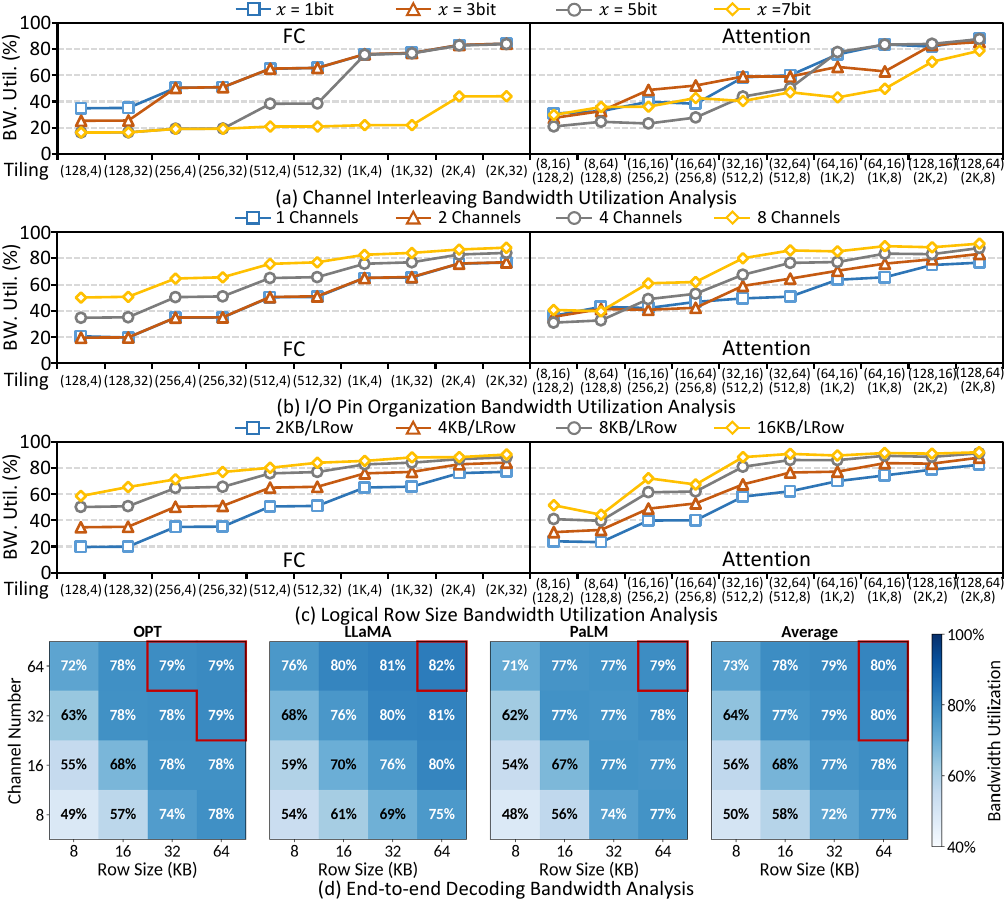}
    \caption{3D-DRAM Organization DSE for Edge LLM.}
    \label{fig:edge-dram-dse}
\end{figure}

%% file: fig_tex/edge-matrix-vector-dse.tex
\begin{figure}
    \centering
    \includegraphics[width=0.99\columnwidth]{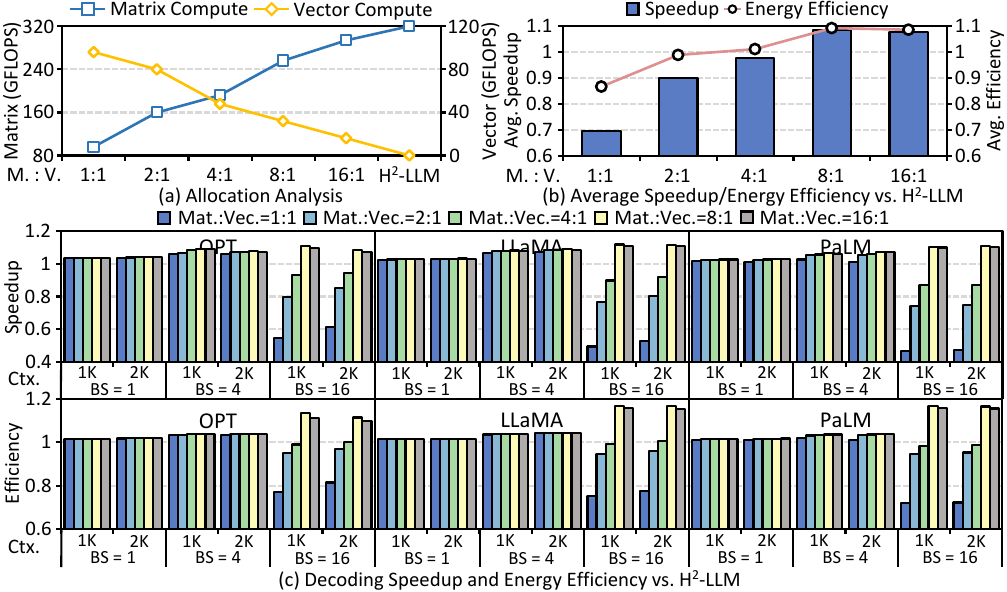}
    \caption{Matrix-Vector Allocation Analysis for Edge LLM.}
     \vspace{-0.5em}
    \label{fig:edge-matrix-vector-dse}
\end{figure}

%% file: tex/8-conclusion.tex
\section{Conclusion}

This paper proposes \TitleAbbr, the first silicon-proven full-stack simulation framework for 3D-Accelerators.
Based on commercially mature 3D-DRAM technology, we develop a general architecture template applicable across diverse scenarios.
We further design hierarchical programming model and corresponding primitives to enhance 3D-Accelerator's useability in real deployments.
Built upon these abstractions, \TitleAbbr~enables cycle-level simulation from architectural configurations and operator implementations, achieving high fidelity to real hardware.
Leveraging \TitleAbbr~for design space exploration, we derive key takeaways for future 3D-Accelerator design and advance the research for 3D-DRAM-based systems.